\documentclass[aps,pre,twocolumn,superscriptaddress,showpacs]{revtex4}
\usepackage{graphicx}
\usepackage[colorlinks=false,linkcolor=blue]{hyperref}
\usepackage[header,title,page,titletoc]{appendix}
\usepackage{epstopdf}
\usepackage{color}
\usepackage{float}
\usepackage{titlesec}

\renewcommand{\thesubsection}{\arabic{subsection}.\hspace{-.4cm}}
\renewcommand{\thesubsubsection}{\arabic{subsection}.\arabic{subsubsection}.\hspace{-.3cm}}

\titleformat{\section}{\flushleft\bfseries\uppercase}{}{1em}{}
\titlespacing{\section}{0pt}{3mm}{3mm}
\titleformat{\subsection}{\flushleft\bfseries}{\thesubsection}{1em}{}
\titlespacing{\subsection}{0pt}{2.5mm}{2.5mm}
\titleformat{\subsubsection}{\flushleft\bfseries}{\thesubsubsection}{1em}{}
\titlespacing{\subsubsection}{0pt}{2.5mm}{2.5mm}

\begin{document}


\title{A critical survey on the kinetic assays of DNA polymerase fidelity from a new theoretical perspective}

\author{Qiu-Shi Li}
 \affiliation{School of Physical Science, University of Chinese Academy of Sciences}%

\author{Yao-Gen Shu}%
\email{shuyaogen@wiucas.ac.cn}
\affiliation{Wenzhou Institute, University of Chinese Academy of Sciences}%

\author{Zhong-Can Ou-Yang}%
\affiliation{Institute of Theoretical Physics, Chinese Academy of Sciences}%

\author{Ming Li}
\email{liming@ucas.ac.cn}
\affiliation{School of Physical Science, University of Chinese Academy of Sciences}%

\date{\today}

\begin{abstract}
The high fidelity of DNA polymerase is critical for the faithful replication of genomic DNA. Several approaches were proposed to quantify the fidelity of DNA polymerase. Direct measurements of the error frequency of the replication products definitely give the true fidelity but turn out very hard to implement. Two biochemical kinetic approaches, the steady-state assay and the transient-state assay, were then suggested and widely adopted. In these assays, the error frequency is indirectly estimated by using the steady-state or the transient-state kinetic theory combined with the measured kinetic rates. However, whether these indirectly estimated fidelities are equivalent to the true fidelity has never been clarified theoretically, and in particular there are different strategies to quantify the proofreading efficiency of DNAP but often lead to inconsistent results. The reason for all these confusions is that it's mathematically challenging to formulate a rigorous and general theory of the true fidelity. Recently we have succeeded to establish such a theoretical framework. In this paper, we develop this theory to make a comprehensive examination on the theoretical foundation of the kinetic assays and the relation between fidelities obtained by different methods. We conclude that while the steady-state assay and the transient-state assay can always measure the true fidelity of exonuclease-deficient DNA polymerases, they only do so for exonuclease-efficient DNA polymerases conditionally (the proper way to use these assays to quantify the proofreading efficiency is also suggested).  We thus propose a new kinetic approach, the single-molecule assay, which indirectly but precisely characterizes the true fidelity of either exonuclease-deficient or exonuclease-efficient DNA polymerases.
\end{abstract}


\maketitle

\section*{Introduction}\label{Introduction}
The high fidelity of DNA polymerase (DNAP) is critical for faithful replication of genomic DNA. Quantitative studies on DNAP fidelity began in 1960s¡¯ and became an important issue in biochemistry and molecular biology. Intuitively, the DNAP fidelity can be roughly understood as the reciprocal of the overall mismatch (error) frequency when a given DNA template is replicated with both the matched dNTPs (denoted as dRTP or R) and the mismatched dNTPs (denoted as dWTP or W). For instance, the synthetic polymer poly-A$\overline{BU}$ was used as the template and the replication reaction was conducted with both dRTPs (dATP and d$\overline{BU}TP$) and dWTP (dGTP). The ratio of the incorporated dRTPs to dWTPs in the final products was then determined to quantify the overall error frequency\cite{Trautner1962}. Similarly, a homopolymer poly-dC was used as the template and the total number of the incorporated dWTP (dTTP) and dRTP (dGTP) was then measured to give the error frequency\cite{Hall1968}. Beyond such overall fidelity, the site-specific fidelity was defined as the reciprocal of the error frequency at individual template sites.  In principle, the error frequency at any template site can be directly counted if a sufficient amount of full-length replication products can be collected and sequenced(this will be denoted as true fidelity $f_{ture}$), \textit{e.g.} by using deep sequencing techniques \cite{Lee2016,DePaz2018}.
However, this type of sequencing-based approach always requires a huge workload and was rarely adopted in fidelity assay. It is also hard to specify the sequence-context influences on the fidelity. A similar but much simpler strategy is to only investigate the error frequency at the assigned template site by single-nucleotide incorporation assays. Such assays are conducted for $exo^-$-DNAP (exonuclease-deficient DNAP), in which dRTP and dWTP compete to be incorporated to the primer terminal only at the assigned single template site and the amount of the final reaction products containing the incorporated dRTP or dWTP are then determined by gel analysis to give the error frequency, $e.g.$\cite{Clayton1979,Bertram2010}.
By designing various template sequences, one can further dissect the sequence-context dependence of the site-specific error frequency.
Although the above definitions of DNAP fidelity are simple and intuitive, the direct measurements are very challenging since mismatches occur with too low frequency to be detected even when heavily-biased dNTP pools are used. Besides, the single-nucleotide incorporation assays do not apply to $exo^+$-DNAP (exonuclease-deficient DNAP) because the coexistence of the polymerase activity and the exonuclease activity makes the reaction products very complicated and hard to interpret. Hence two alternative kinetic approaches were proposed.

The steady-state method was developed by A. Fersht for $exo^-$-DNAP, which is based on the Michaelis-Menten kinetics of the incorporation of a single dRTP or dWTP at the same assigned template site\cite{Fersht1985}.The two incorporation reactions are conducted separately under steady-state conditions to obtain the specificity constant (the quasi-first order rate constant) $(k_{cat}/K_{m})_{R}$ or $(k_{cat}/K_{m})_{W}$ respectively, $k_{cat}$ is the maximal steady-state turnover rate of dNTP incorporation and $K_{m}$ is the Michaelis constant. The site-specific fidelity is then characterized as the ratio between the two incorporation velocities, \textit{i.e.} $(k_{cat}/K_{m})_{R}\textmd{[dRTP]}\big/(k_{cat}/K_{m})_{W}\textmd{[dWTP]}$ (denoted as steady-state fidelity $f_{s\cdot s}$) , which is nothing but the specificity commonly defined for multi-substrate enzymes.  This assay has been widely acknowledged as the standard method in DNAP fidelity studies. Nevertheless, there is an apparent difference between the specificity and the true fidelity of $exo^-$-DNAP. Enzyme specificity is operationally defined and measured under the steady-state condition which is usually established in experiments by two requirements, \textit{i.e.} the substrate is in large excess to the enzyme, and the enzyme can dissociate from the product after a single turnover is finished. These two requirements are often met by many reactions catalyzed by non-processive enzymes, and the enzyme specificity is indeed a good measure of the relative contents of final products of competing substrates. DNAP, however, is a processive enzyme and rarely dissociates from the template, which violates the second requirement. Additionally, DNA replication \textit{in vivo} consists of only a single template DNA but many DNAPs, which violates the first requirement. Hence, no steady-state assumptions can be made \textit{a priori} to single-nucleotide incorporation reactions either $in\ vivo$ or $in\ vitro$.  So, is the enzyme specificity really relevant to the true fidelity of $exo^-$-DNAP ?  So far as we know, there was only one experiment work which did the comparison and indicated the possible equivalence of $f_{s\cdot s}$ to $f_{true}$ for Klenow fragment ($\textmd{KF}^-$)\cite{Bertram2010}, but no theoretical works have ever been published to investigate the true fidelity of DNA replication and examine the equivalence of $f_{s\cdot s}$ and $f_{true}$ in general.

Besides the steady-state method, the transient-state kinetic analysis was also proposed to obtain the specificity constant\cite{Kati1992,Johnson1993}. Under the pre-steady-state condition or the single-turnover condition,  one can obtain the parameter $k_{pol}/K_{d}$ (a substitute for $k_{cat}/K_{m}$) for the single-nucleotide incorporation reactions with $exo^-$-DNAP, and define the site-specific fidelity as $(k_{pol}/K_{d})_{R}\textmd{[dRTP]}\big/(k_{pol}/K_{d})_{W}\textmd{[dWTP]}$ (denoted as transient-state fidelity $ f_{t\cdot s}$).  Either $k_{pol}/K_d$ or $k_{cat}/K_{m}$ can only be properly interpreted by kinetic models, so the relation between the two parameters is actually model-dependent. For the commonly used two-step kinetic model (including only dNTP binding and the subsequent chemical step),  it can be shown that they are equal\cite{Johnson1992}. For complex models including additional steps ($e.g.$ DNA binding to DNAP, translocation of DNAP on the template, PPi release, etc.),  their equivalence can also be proved in general (details will given in later sections). But again the relevance of $f_{t\cdot s}$ to $f_{true}$ is not yet clarified.  Although the experiment has indicated the possible equivalence of $f_{t\cdot s}$ and $f_{true}$ for $\textmd{KF}^-$\cite{Bertram2010}, a general theoretical examination is still needed.

Further, these methods fail to definitely measure the site-specific fidelity of $exo^+$-DNAP. For $exo^+$-DNAP, the total fidelity is often assumed to consist of two multiplier factors. The first is the initial discrimination $f_{ini}$ contributed solely by the polymerase domain, which can be given by $f_{s\cdot s}$ or $f_{t\cdot s}$. The second factor is the additional proofreading efficiency $f_{pro}$ contributed by the exonuclease domain, which is defined by the ratio of the elongation probability of the terminal R ($P_{el,R}$) to that of the terminal W ($P_{el,W}$).  Here the elongation probability is given by $P_{el} = k_{el}/ (k_{el} + k_{ex})$,  $k_{el}$ is the elongation rate to the next site, and $k_{ex}$ is the excision rate of the terminal nucleotide at the assigned site ($e.g.$ Eq.(A1-A6) in Ref.\cite{Fersht1982}). $P_{el,R}$ is usually assumed close to 100\%, so $f_{pro}$ equals  approximately to $1+ k_{ex,W}/ k_{el,W}$.  Although these expressions seem reasonable, there are some problems that were not clarified. First, the definition of $f_{pro}$ is subjective though intuitive, so a rigorous theoretical foundation is needed. Second, the rate parameters $ k_{el}$ and $k_{ex}$ are not well defined since both the elongation and the excision are multi-step processes, \textit{\textit{i.e.}} $k_{el}$ and $k_{ex}$ are unknown functions of the involved rate constants but there is not a unique way to define them. They could be theoretically defined under steady-state assumptions (Eq.(6) in Ref.\cite{Fersht1979}) or operationally defined by experiment assays ($e.g.$ steady-state assays\cite{Wingert2013,Vashishtha2015} or transient-state assays \cite{Donlin1991}), but different ways often lead to inconsistent interpretations and quite different estimates of $f_{pro}$ (as will be clarified in  \uppercase{Results and discussion} Sec.2).  Additionally, $k_{el}$ should be more properly understood as the effective elongation rate in the sense that the elongated terminal (the added nucleotide) is no longer excised. This condition is not met if the $exo^+$-DNAP can proofread the buried mismatches ($e.g.$ the penultimate or antepenultimate mismatches, etc.) . In these cases, $k_{el}$ is affected not only by the next template site but also by further sites. Such far-neighbor effects were not seriously considered in previous studies. So, what on earth is the relation between the total fidelity $f _{tot}$ ($=f_{ini}\cdot f_{pro}$) and $f_{true}$?

Recently two equivalent rigorous theories were proposed to investigate the true fidelity of either $exo^-$-DNAP or $exo^+$-DNAP, \textit{i.e.} the iterated function systems by P.Gaspard \cite{Gaspard2017}
and the first-passage (FP) method by us \cite{Li2019}.
In particular, we have obtained very simple and intuitive mathematical formulas by FP method to compute rigorously $f_{true}$ of $exo^+$-DNAP, which can not be achieved by the steady-state or the transient-state analysis.  With these firmly established results, we can address all the above questions in detail.  In the following sections, we will first give a brief review of the FP method and the major conclusions already obtained for simplified kinetic models of DNA replication.  Then we will generalize these conclusions to more realistic kinetic models for $exo^-$-DNAP and $exo^+$-DNAP,  and carefully examine the relations between $f_{s\cdot s}$, $f_{t\cdot s}$ and $f_{true}$.  In particular, the FP analysis makes it possible to take full advantage of single-molecule techniques to investigate the site-specific fidelity, whereas the conventional steady-state or transient-state analysis applies only to ensemble reactions but not to single-molecule processes. Feasible single-molecule assays for either $exo^-$-DNAP or $exo^+$-DNAP will also be suggested.

\section*{Methods}\label{Sec-FP}
\subsection{Basics of the FP method}
The first-passage (FP) method was proposed to study the replication of the entire template by $exo^+$-DNAP \cite{Li2019}, which also applies to single-nucleotide incorporation reactions.
\begin{figure}[H]
\centering
\includegraphics[width=6.5cm]{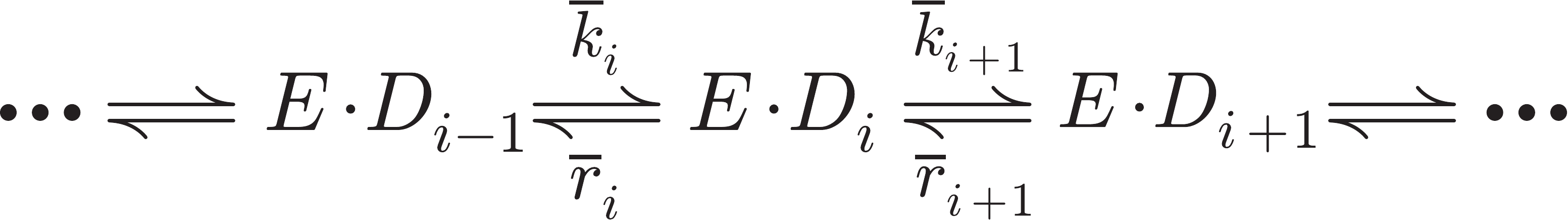}
\caption{The highly simplified reaction scheme of DNA replication. $E$: the enzyme DNAP. $D_i$: the primer-template duplex with primer terminal at the template site $i$.}
\label{Fig_FP_single_step}
\end{figure}
Here the highly simplified reaction scheme Fig.\ref{Fig_FP_single_step} is taken as an example to illustrate the basic logic of this method.  $\overline{k}_i$ is the incorporation rate of dNTP to the primer terminal at the template site $i-1$ (the dNTP-concentration dependence of $\overline{k}_i$ is not explicitly shown here),   $\overline{r}_i$ is the excision rate of the primer terminal at the template site $i$.  In Fig.\ref{Fig_FP_single_step}, dRTP and dWTP compete for each template site during the replication, so there will be various sequences in the final full-length products. The FP method describes the entire template-directed replication process by chemical kinetic equations, and directly compute the sequence distribution of the full-length products from which $f_{true}$ can be precisely calculated. It is worth noting that the FP method does not need any extra assumptions like steady-state or quasi-equilibrium assumptions, or need to explicitly solve the kinetic equations as done in the transient-state analysis which is often a formidable task. Some illustrative examples of FP calculations will be given in later sections. Here we only list the major results in terms of  $\overline{k}_i$ and $\overline{r}_i$.  Detailed computation can be found in Ref.\cite{Li2019}

Intuitively, $\overline{k}_i$ and $\overline{r}_i$ depend on the identity (A, G, T or C) and the state (matched or mismatched) not only of the base pair at site $i$ but also of the one or more preceding base pairs.  If there are only nearest-neighbor (first-order) effects,  $\overline{k}_i$ and $\overline{r}_i$ can be written as  ${\overline{k}}^{X_{i-1}X_{i}}_{\alpha_{i-1}\alpha_{i}}$ and ${\overline{r}}^{X_{i-1}X_{i}}_{\alpha_{i-1}\alpha_{i}}$, $X_{i-1}$ (or $X_{i}$) represents the nucleotide at site $i-1$ (or $i$) on the template,  $\alpha_{i-1}$ represents the nucleotide at site $i-1$ on the primer, $\alpha_{i}$ represents the the next nucleotide to be incorporated to the primer terminal at site $i$ (for ${\overline{k}}^{X_{i-1}X_{i}}_{\alpha_{i-1}\alpha_{i}}$) or the terminal nucleotide of the primer at site $i$ to be excised (for ${\overline{r}}^{X_{i-1}X_{i}}_{\alpha_{i-1}\alpha_{i}}$).  $X$ and $\alpha$ can be any of the four types of nucleotides A, G, T and C.  Similarly,  there are ${\overline{k}}^{X_{i-2}X_{i-1}X_{i}}_{\alpha_{i-2}\alpha_{i-1}\alpha_{i}}$ etc. for the second-order neighbor effects, and so on for far-neighbor (higher-order) effects.

\subsection{The true fidelity calculated by the FP method}

For DNAP having first-order neighbor effects, in a wide range of the involved rate constants, we have derived the analytical expression of the fidelity at site $i$ \cite{Li2019},
\begin{eqnarray}\label{Fid-eq-muti}
f_{true,i}\approx
\left[\sum\limits_{W_{i}\neq R_{i}}\frac{\overline{k}^{X_{i-1}X_{i}}_{R_{i-1}W_{i}}}{{\overline{k}}^{X_{i-1}X_{i}}_{R_{i-1}R_{i}}}
\frac{{\overline{k}}^{X_{i}X_{i+1}}_{W_{i}R_{i+1}}}{{\overline{k}}^{X_{i}X_{i+1}}_{W_{i}R_{i+1}}
+{\overline{r}}^{X_{i-1}X_{i}}_{R_{i-1}W_{i}}}\right]^{-1}
\end{eqnarray}
$R$ represents the matched nucleotide, and $W$ represents any one of the three types of mismatched nucleotides.  For simplicity, we omit all the superscripts below unless it causes misunderstanding. Each term in the sum represents the error frequency of a particular type of mismatch, whose reciprocal is the mismatch-specific fidelity studied in the conventional steady-state assay or transient-state assay,
\begin{eqnarray}\label{Fid-eq}
f_{true,i}\approx f^{pol}_{i}\cdot f^{exo}_{i}
\end{eqnarray}
where
\begin{eqnarray}\label{Fid-eq-pol}
f^{pol}_{i}\approx
 \frac{\overline{k}_{R_{i-1}R_{i}}}{\overline{k}_{R_{i-1}W_{i}}}
\end{eqnarray}
is the initial discrimination, and
\begin{eqnarray}\label{Fid-eq-exo}
f^{exo}_{i}\approx
1+\frac{\overline{r}_{R_{i-1}W_{i}}}{\overline{k}_{W_{i}R_{i+1}}}
\end{eqnarray}
is the proofreading efficiency. This is similar to $f_{pro}$ defined in  \uppercase{Introduction}, if $\overline{k}_{W_{i}R_{i+1}},\overline{r}_{R_{i-1}W_{i}}$ are regarded as $k_{el,W}$ $k_{ex,W}$ respectively.

For DNAP having second-order neighbor effects, with some reasonable assumptions about the rate parameters, we can obtain the fidelity at site $i$ \cite{Li2019},
\begin{eqnarray}\label{Fid_eq_high-order}
&&f_{true,i}\approx\nonumber\\
&&\Big[\sum\limits_{W_{i}\neq R_{i}}
\frac{\overline{k}_{R_{i-2}R_{i-1}W_{i}}}{\overline{k}_{R_{i-2}R_{i-1}R_{i}}}
\frac{k^{el}_{R_{i-1}W_{i}R_{i+1}}}{k^{el}_{R_{i-1}W_{i}R_{i+1}}+\overline{r}_{R_{i-2}R_{i-1}W_{i}}}\Big]^{-1} \nonumber  \\
&&k^{el}_{R_{i-1}W_{i}R_{i+1}}=\frac{\overline{k}_{R_{i-1}W_{i}R_{i+1}}\overline{k}_{W_{i}R_{i+1}R_{i+2}}}
{\overline{k}_{W_{i}R_{i+1}R_{i+2}}+\overline{r}_{R_{i-1}W_{i}R_{i+1}}}
\end{eqnarray}
Each term in the sum represents the mismatch-specific error frequency at site $i$. Its reciprocal defines the mismatch-specific fidelity which again consists of the initial discrimination and the proofreading efficiency, but the latter differs significantly from $f_{pro}$ in  \uppercase{Introduction}, since the effective elongation rate is not $\overline{k}_{R_{i-1}W_{i}R_{i+1}}$ but instead $k^{el}_{R_{i-1}W_{i}R_{i+1}}$ which includes the next-nearest neighbor effects. The same logic can be readily generalized to higher-order neighbor effects where the proofreading efficiency will be more complicated \cite{Song2017,Li2019}.

In real DNA replication, either the dNTP incorporation or the dNMP excision is a multi-step process.  By using the FP method, the complex reaction scheme can be reduced to the simplified scheme Fig.\ref{Fig_FP_single_step}, and the fidelity can still be calculated by Eq.(\ref{Fid-eq})or Eq.(\ref{Fid_eq_high-order}), with only one modification: $\overline{k}$ and $\overline{r}$ are now the effective incorporation rates and the effective excision rates respectively which are functions of the involved rate constants. In the following sections, we will derive these functions for different multi-step reaction models, and compare them with those obtained by steady-state or transient-state assays. For simplicity, we only discuss DNAP having first-order neighbor effects in details, since almost all the existing literature focused on this case. Higher-order neighbor effects will also be mentioned in  \uppercase{Summary}.

\section*{Results and discussion}
\setcounter{subsection}{0}
\subsection{Fidelity assays of exo$^-$-DNAP}\label{Sec-pol}
\subsubsection{The true fidelity measured by the direct competition assay}\label{Sec-pol-direct}

\begin{figure}[H]
\centering
\includegraphics[width=7cm]{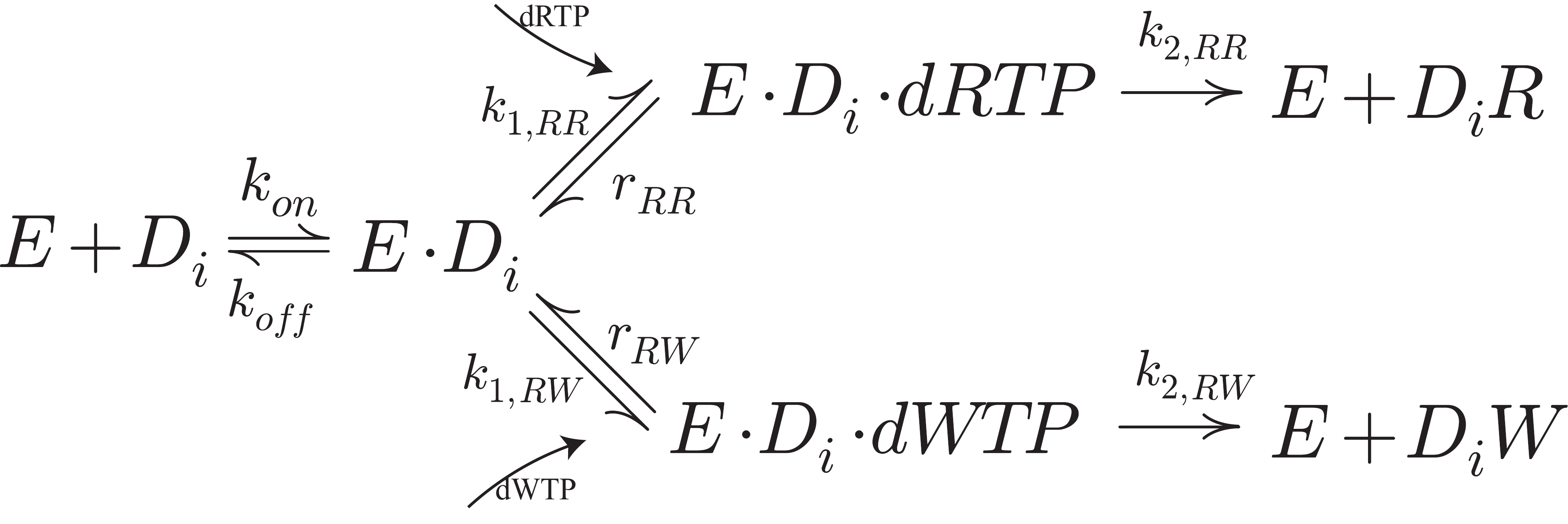}
\caption{The minimal reaction scheme of the competitive incorporation of dRTP and dWTP. $E$: $exo^-$-DNAP. $D_i$: the primer-template duplex with the matched(R) terminal at site $i$. For brevity, the subscript $i$ in each rate constant is omitted.}
\label{Fig_direct}
\end{figure}
Fig.\ref{Fig_direct} shows a three-step kinetic model of the competitive incorporation of a single dRTP or dWTP to site $i+1$. The true fidelity is precisely given by the ratio of the final product $D_{i}R$ to $D_{i}W$ when the substrate DNA are totally consumed, \textit{i.e.},
\begin{eqnarray}\label{direct-ture-fidelity}
f_{ture}=\frac{[D_{i}R]}{[D_{i}W]}
\end{eqnarray}
which can be calculated by FP method.

A part of the kinetic equations for this model are given below ,
\begin{eqnarray}\label{direct-product-all}
\frac{d}{dt}[D_{i}\alpha]&=&k_{2,R\alpha}[E\cdot D_{i}\cdot d\alpha TP]\nonumber\\
\frac{d}{dt}[E\cdot D_{i}\cdot d\alpha TP]&=&k^{0}_{1,R\alpha}[E\cdot D_{i}][d\alpha TP]\nonumber\\
&-&(k_{2,R\alpha}+r_{R\alpha})[E\cdot D_{i}\cdot d\alpha TP]\nonumber\\
\end{eqnarray}
here $\alpha=$R,W. The dNTP binding rate is denoted as $k_{1,R\alpha}=k^{0}_{1,R\alpha}[d\alpha TP]$. The basic idea of FP method is not to directly solve the kinetic equations rigorously ($e.g.$ in the transient-state analysis) or approximately by imposing extra assumptions ($e.g.$ the steady-state assumption) . Instead, the two equations are integrated to give the products at time $t$,
\begin{eqnarray}\label{direct-product}
[D_{i}\alpha](t)&=&\frac{k^{0}_{1,R\alpha}k_{2,R\alpha}}{k_{2,R\alpha}+r_{R\alpha}}\int^{t}_{0}([E\cdot D_{i}](\tau)[d\alpha TP](\tau))d\tau\nonumber \\
&-&\frac{k_{2,R\alpha}}{k_{2,R\alpha}+r_{R\alpha}}[E\cdot D_{i}\cdot d\alpha TP](t)
\end{eqnarray}
The second term approaches to zero with $t$ increases to infinity.  dNTP is usually in large excess to template DNA either $in\ vivo$ or $in\ vitro$, so $\textmd{[dNTP]}$ remains approximately a constant during the reaction. Then the fidelity is simply given by
\begin{eqnarray}\label{direct-fidelity2}
f_{ture}&=&\overline{k}_{RR}\Big/\overline{k}_{RW}\nonumber\\
\overline{k}_{RR}&=&\frac{k^{0}_{1,RR}k_{2,RR}}{k_{2,RR}+r_{RR}}\textmd{[dRTP]}\nonumber\\
\overline{k}_{RW}&=&\frac{k^{0}_{1,RW}k_{2,RW}}{k_{2,RW}+r_{RW}}\textmd{[dWTP]}
\end{eqnarray}
$f_{true}$ is exactly the initial discrimination defined by Eq.(\ref{Fid-eq-pol}) with the two effective incorporation rates $ \overline{k}_{RR}$ and $\overline{k}_{RW}$.

In practice, when the reaction time $t$ is large enough for sufficient product accumulation (\textit{i.e.}, the second term on the right side of Eq.(\ref{direct-product}) is far smaller than the first term), the measured $[D_{i}R](t)/[D_{i}W](t)$ becomes nearly time-invariant, and thus it is a good measure of $f_{true}$.
In the direct competition assay conducted by Bertram \textit{et.al}\cite{Bertram2010}, the incorporation reaction was terminated when about half of the substrate DNA were reacted. This termination criteria \textit{per se} does not meet the above requirement. Other evidences should be considered. For instance, $[D_{i}R](t)/[D_{i}W](t)$ is proportional to $\textmd{[dRTP]}/\textmd{[dWTP]}$ if the reaction time $t$ is large, so one can decide whether $t$ is sufficient large by examining whether $[D_{i}R](t)\textmd{[dWTP]}/[D_{i}W](t)\textmd{[dRTP]}$ becomes nearly a constant when $\textmd{[dWTP]}$ or $\textmd{[dRTP]}$ is changed.
Combined with these evidences, Bertram \textit{et.al} were able to show that $[D_{i}R](t)/[D_{i}W](t)$ measured under their termination condition is really a good measure of the true fidelity.

\subsubsection{Effective rates of multi-step reactions uniquely determined by FP method}\label{Sec-pol-FP}
\begin{figure}[H]
\centering
\includegraphics[width=7.5cm]{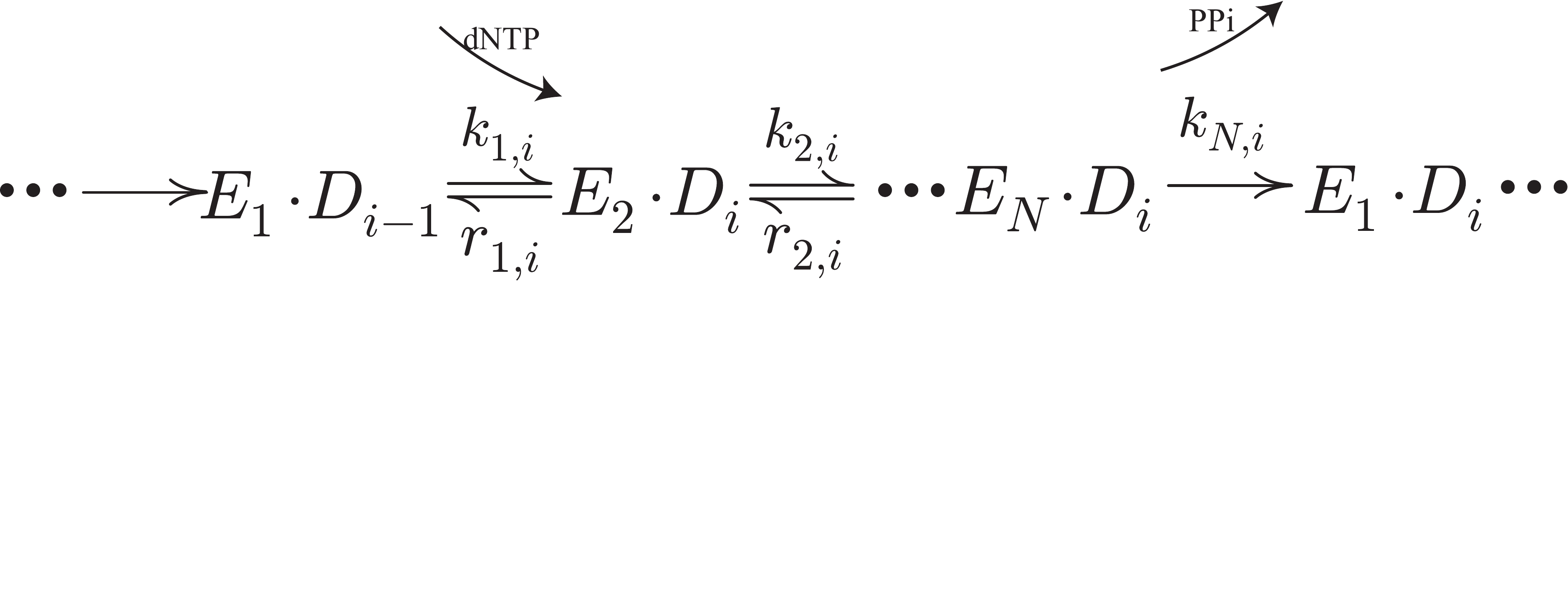}
\vspace{-1.5cm}
\caption{The multi-step incorporation scheme. The enzyme-substrate complex ($E\cdot D$) goes through $N$ states (indicated by subscripts $1,...,N$) to successfully incorporate a single nucleotide(indicated by subscript $i$). To simplify the notation of the rate constants, the superscripts indicating the template nucleotide $X_i$ and the subscripts indicating the primer nucleotide $\alpha_i$ are omitted. This omission rule also applies to other figures in this paper, unless otherwise specified.}
\label{Fig_Pol_nt_FP}
\end{figure}

The above FP treatment can be directly extended to multi-state incorporation schemes like Fig.\ref{Fig_Pol_nt_FP} to get the effective incorporation rate, as below,
\begin{eqnarray}\label{Pol-nt-FP-ieq1}
&\overline{k}=k^{*}=k^{'}_{N-1}k_{N}\big/(r^{'}_{N-1}+k_{N})&\nonumber\\
&k^{'}_j=k^{'}_{j-1}k_{j}\big/(r^{'}_{j-1}+k_{j})&\nonumber\\
&r^{'}_j=r^{'}_{j-1}r_{j}\big/(r^{'}_{j-1}+k_{j})&\nonumber\\
&(j=N,N-1,...,3)&\nonumber\\
&k^{'}_2=k_{1}k_{2}\big/(r_{1}+k_{2})&\nonumber\\
&r^{'}_2=r_{1}r_{2}\big/(r_{1}+k_{2})&
\end{eqnarray}
Details of the calculation can be found in Supplementary Materials (SM) Sec.\uppercase\expandafter{\romannumeral 1} B.
Here $k_{1}$ is proportional to dNTP concentration $k_{1}=k^0_{1}\textmd{[dNTP]}$, so $k^*=k^{*0}\textmd{[dNTP]}$. The true fidelity is still given by Eq.(\ref{Fid-eq-pol}) where $\overline{k}_{RR}$ and $\overline{k}_{RW}$ are effective rates defined here.
Fig.\ref{Fig_Pol_nt_FP} describes the processive dNTP incorporation by DNAP without dissociation from the substrate DNA.  If the dissociation is considered, the reaction scheme will be more complex, but the effective incorporation rates can still be given as above, as will be shown in  \uppercase{Results and discussion} Sec.2.2.

\subsubsection{The steady-state assay measures the true fidelity}\label{Sec-pol-ss}
The steady-state assays measure the initial velocity of product generation under the condition that the substrate is in large excess to the enzyme. The normalized velocity per enzyme is in general given by the Michaelis-Menten equation
\begin{eqnarray}\label{Pol-nt-ss-specificity1}
v^{pol}_{s\cdot s}=\frac{k_{cat}\textmd{[dNTP]}}{\textmd{[dNTP]}+K_m}
\end{eqnarray}
Here the superscript $pol$ indicates the polymerase activity, the subscript $s\cdot s$ indicates the steady state. Fitting the experimental data by this equation, one can get the specific constant $k_{cat}/K_m$ either for dRTP incorporation or dWTP incorporation and estimate the fidelity (the initial discrimination) by $f_{s\cdot s}=(k_{cat}/K_{m})_{R}\textmd{[dRTP]}\big/(k_{cat}/K_{m})_{W}\textmd{[dWTP]}$.
What is the relation between $k_{cat}/K_m$ and the effective incorporation rate $\overline{k}$ in Eq.(\ref{Fid-eq-pol})?

\begin{figure}[H]
\centering
\includegraphics[width=6.5cm]{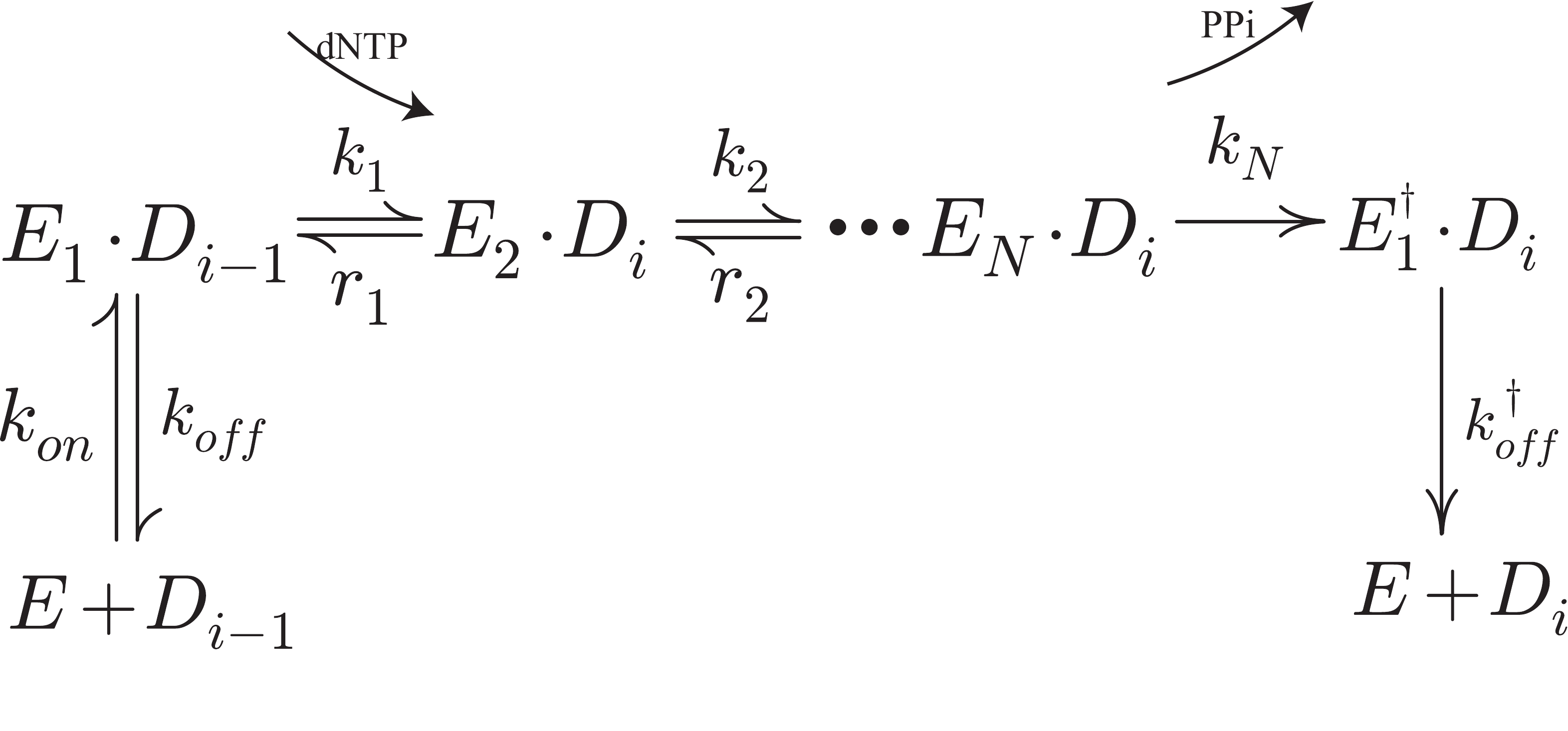}
\caption{The reaction scheme for the steady-state assay to measure the specificity constant of the nucleotide incorporation reaction of $exo^-$DNAP.}
\label{Fig_Pol_nt_ss}
\end{figure}
To understand the exact meaning of $k_{cat}/K_m$, the complete multi-step incorporation reaction scheme Fig.(\ref{Fig_Pol_nt_ss}) must be considered, which explicitly includes the DNAP binding step and the dissociation step. The last dissociation step is reasonably assumed irreversible, since the enzyme will much unlikely rebind to the same substrate molecule after dissociation because the substrate is in large excess to the enzyme. Under the steady-state condition, it can be easily shown
\begin{eqnarray}\label{Pol-nt-ss-specificity2}
\frac{k_{cat}\textmd{[dNTP]}}{K_m}=\frac{k^*}{K_{s\cdot s}}
\end{eqnarray}
Here $k^*$ is defined in Eq.(\ref{Pol-nt-FP-ieq1}). $K_{s\cdot s}=1+k_{off}/k_{on}$, $k_{on}=k^{0}_{on}\textmd{[DNA]}$. $K_{s\cdot s}$ is exactly the same for either dRTP incorporation or dWTP incorporation. So the fidelity is given as
\begin{eqnarray}\label{Pol-nt-ss-fidelity}
f_{s\cdot s}\equiv\frac{(k_{cat}\textmd{[dNTP]}/K_m)_{R_{i-1}R_{i}}}{(k_{cat}\textmd{[dNTP]}/K_m)_{R_{i-1}W_{i}}}
=\frac{k^*_{R_{i-1}R_{i}}}{k^*_{R_{i-1}W_{i}}}=f_{ture}\nonumber\\
\end{eqnarray}
This is understandable: the steps before dNTP binding should not contribute to the initial discrimination. However, it does not mean that those steps do not contribute to the total fidelity.  Actually they can affect the proofreading efficiency, as will be demonstrated in  \uppercase{Results and discussion} Sec.2.

\subsubsection{The transient-state assay measures the true fidelity}\label{Sec-pol-ts}
The transient-state assay often refers to two different methods, the pre-steady-state assay or the single-turnover assay. Since the theoretical foundations of these two methods are the same, we only discuss the latter below for simplicity.

In single-turnover assays, the enzyme is in large excess to the substrate, and so the dissociation of the enzyme from the product is neglected. The time course of the product accumulation or the substrate consumption is monitored. The data is then fitted by exponential functions (single-exponential or multi-exponential) to give one or more exponents (\textit{i.e.} the characteristic rates). In DNAP fidelity assay, these rates are complex functions of all the involved rate constants and dNTP concentration, which in principle can be analytically derived for any given kinetic model. For instance, for the commonly-used simple model including only substrate binding and the subsequent irreversible chemical step, one can directly solve the kinetic equations to get two rate functions. It was proved by K. Johnson that the smaller one obeys approximately the Michaelis-Menten-like equations\cite{Johnson1992}.
\begin{eqnarray}\label{Pol-nt-ts-specificity}
v^{pol}_{t\cdot s}=\frac{k_{pol}\textmd{[dNTP]}}{\textmd{[dNTP]}+K_d}
\end{eqnarray}
The subscript $t\cdot s$ indicates the transient state. Similar to the steady-state assays, $k_{pol}/K_d$ is regarded as the specific constant and thus DNAP fidelity is defined as the enzyme specificity $ f_{t\cdot s}=(k_{pol}/K_{d})_R\textmd{[dRTP]}/(k_{pol}/K_{d})_W\textmd{[dWTP]}$.  It was also shown that $k_{pol}/K_d$ equals to $k_{cat}/K_m$ for the two-step model\cite{Johnson1992}, so $f_{t\cdot s} = f_{s\cdot s}$.

\begin{figure}[H]
\centering
\includegraphics[width=7.3cm]{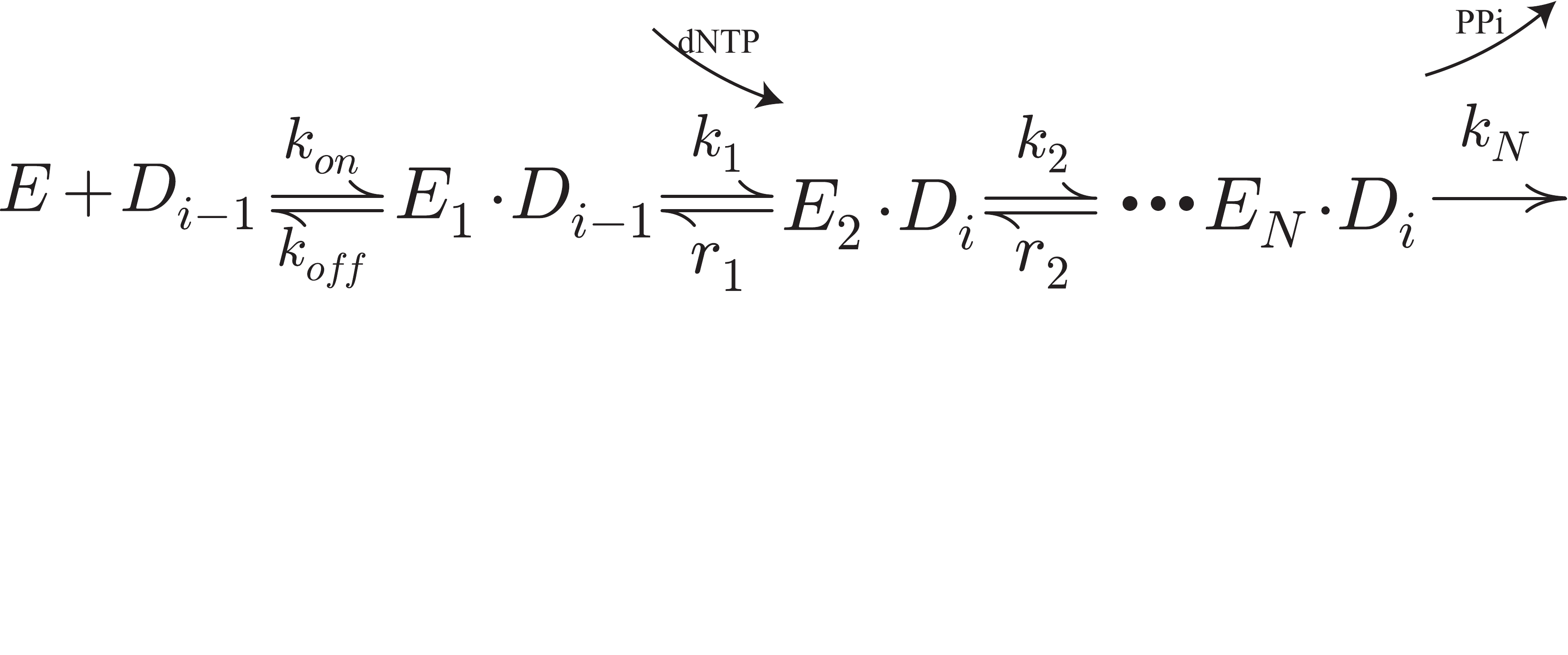}
\vspace{-1.5cm}
\caption{The reaction scheme for the transient-state assay to measure the specificity constant of the nucleotide incorporation reaction of $exo^-$-DNAP.}
\label{Fig_Pol_nt_ts}
\end{figure}
The equality $k_{pol}/K_d=k_{cat}/K_m$ actually holds for more general models like Fig.\ref{Fig_Pol_nt_ts}. The rigorous proof is too lengthy to be presented here (details can be found in SM Sec.\uppercase\expandafter{\romannumeral 3} B). Below we only give some intuitive explanations.

Since there are $N$ states in the reaction scheme Fig.\ref{Fig_Pol_nt_ts}, the time evolution of the system can be described by $N$ exponentially-decay functions with $N$ characteristic rates. If the smallest rate $v^{pol}_{t\cdot s}$ is much smaller than the others, it can be easily proven that $v^{pol}_{t\cdot s}$  follows the same form as Eq.(\ref{Pol-nt-ts-specificity}) in general.  Thus $k_{pol}/K_d$ can be obtained by mathematically extrapolating $\textmd{[dNTP]}$ to zero, $v^{pol}_{t\cdot s}\approx k_{pol}\textmd{[dNTP]}/K_d$. Intuitively, when $\textmd{[dNTP]}$ approaches to zero, dNTP binding is the rate-limiting step, and all the steps after it will be so slow that the accumulation of each intermediate state is almost zero, \textit{i.e.} they are approximately in steady state.  This gives the velocity per substrate DNA, $v^{pol}_{t\cdot s}\approx k^*[E_{1}\cdot D_{i-1}]/[D_0]$. $k^*$ is defined by Eq.(\ref{Pol-nt-FP-ieq1}), $[D_0]$ is the total concentration of DNA. On the other hand, all the steps before dNTP binding are relatively much faster and approximately in equilibrium, which leads to $ [E_{1}\cdot D_{i-1}]\approx(k_{on}/k_{off})[D_{i-1}]$, \textit{i.e.} $[E_{1}\cdot D_{i-1}]\approx[D_0]/(1+k_{off}/k_{on})$. Here $k_{on}=k^{0}_{on}[E]$, $[E]$ equals almost to the total DNAP concentration $[E_{0}]$ since DNAP is in large excess to DNA. So the normalized velocity per substrate is $v^{pol}_{t\cdot s}\approx k^*/K_{t\cdot s}$, which leads to
\begin{eqnarray}\label{Pol-nt-ts-specificity2}
\frac{k_{pol}\textmd{[dNTP]}}{K_d}=\frac{k^{*}}{K_{t\cdot s}}
\end{eqnarray}
$K_{t\cdot s}=1+k_{off}/k_{on}$. This is exactly the same as Eq.(\ref{Pol-nt-ss-specificity2}).
So the fidelity can be given by
\begin{eqnarray}\label{Pol-nt-ts-fidelity}
f_{t\cdot s}\equiv\frac{(k_{pol}\textmd{[dNTP]}/K_d)_{R_{i-1}R_{i}}}{(k_{pol}\textmd{[dNTP]}/K_d)_{R_{i-1}W_{i}}}
=\frac{k^*_{R_{i-1}R_{i}}}{k^*_{R_{i-1}W_{i}}}=f_{ture}\nonumber\\
\end{eqnarray}

\subsubsection{A new approach to measure the true fidelity: a single-molecule assay}\label{Sec-pol-sm}
As stated above, neither the steady-state assay nor the transient-state assay can give the effective incorporation rates. This is not a problem for fidelity assay of $exo^-$-DNAP, but is indeed a serious problem for $exo^+$-DNAP (as shown in later sections). Then, how can one estimate the effective rates by a general method ? A possible way is to directly dissect the reaction mechanism, \textit{i.e.} measuring the rate constants of each step by transient-state experiments \cite{Wong1991,Tsai2006,Purohit2003,Joyce2008,Santoso2010,Hohlbein2013}, and then one can calculate the effective rate according to Eq.(\ref{Pol-nt-FP-ieq1}).  This is a perfect approach but needs heavy work.  Are there direct measurements of the effective rates? Here we suggest a possible single-molecule approach based on the FP analysis.

In a typical single-molecule experiment, the different states of the enzyme or the substrate can be distinguished by techniques such as smFRET\cite{Santoso2010,Hohlbein2013}.  So, if the state $E_1\cdot D_{i-1}$ and $E^{\dag}_1\cdot D_{i}$ in Fig.\ref{Fig_Pol_nt_ss} can be properly identified, the following single-molecule experiment can be done to measure the effective incorporation rates.

1. Initiate the nucleotide incorporation reaction by adding $exo^-$-DNAP and dNTP to the substrate $D_{i-1}$ and begin to record the state-switching trajectory of a single enzyme-DNA complex.  Here dNTP can be dRTP or dWTP, and the primer terminal can be matched(R) or mismatched(W). When a single DNAP is captured by the substrate DNA, it can catalyze the incorporation of one or more nucleotides, depending on the template sequence context and the dNTP used. Then one can select a particular time window from the recorded trajectory, starting from the first-arrival at $E_1\cdot D_{i-1}$ (denoted as starting point) and ending at the first-arrival at $E_{1}\cdot D_{i}$ (denoted as ending point).

2. In this time window, the system may make multiple visits to $E_1\cdot D_{i-1}$.  Count the total time the system resides at $E_1\cdot D_{i-1}$. This so-called residence time may be clearly measured under low concentrations of dNTP.

3. Collect sufficient samples to get the averaged residence time  $\Gamma_{1,i-1}$, which gives directly the required effective incorporation rate $k^{*}_{\alpha_{i-1}\alpha_i}=1/{\Gamma_{1}}_{\alpha_{i-1}}$. Here $k^{*}_{\alpha_{i-1}\alpha_i}=k^{*0}_{\alpha_{i-1}\alpha_i}[d\alpha_i TP]$, $\alpha=$A,T,G,C. The proof of this equality is given in SM Sec.\uppercase\expandafter{\romannumeral 4} A.

The advantage of this single-molecule analysis is its model-independence. Since $k^{*}_{i}=1/\Gamma_{1,i-1}$ holds in general, the measurement of  $\Gamma_{1,i-1}$ does not depend on any hypothesis about the details of the reaction scheme (in fact, steps after dNTP binding are often unclear).  So this method is hopefully an alternative of the conventional ensemble assays, particularly in cases where the latter may fail (see later sections).

\subsection{Fidelity assays of exo$^+$-DNAP}\label{Sec-exo}
\begin{figure}[H]
\centering
\includegraphics[width=7.3cm]{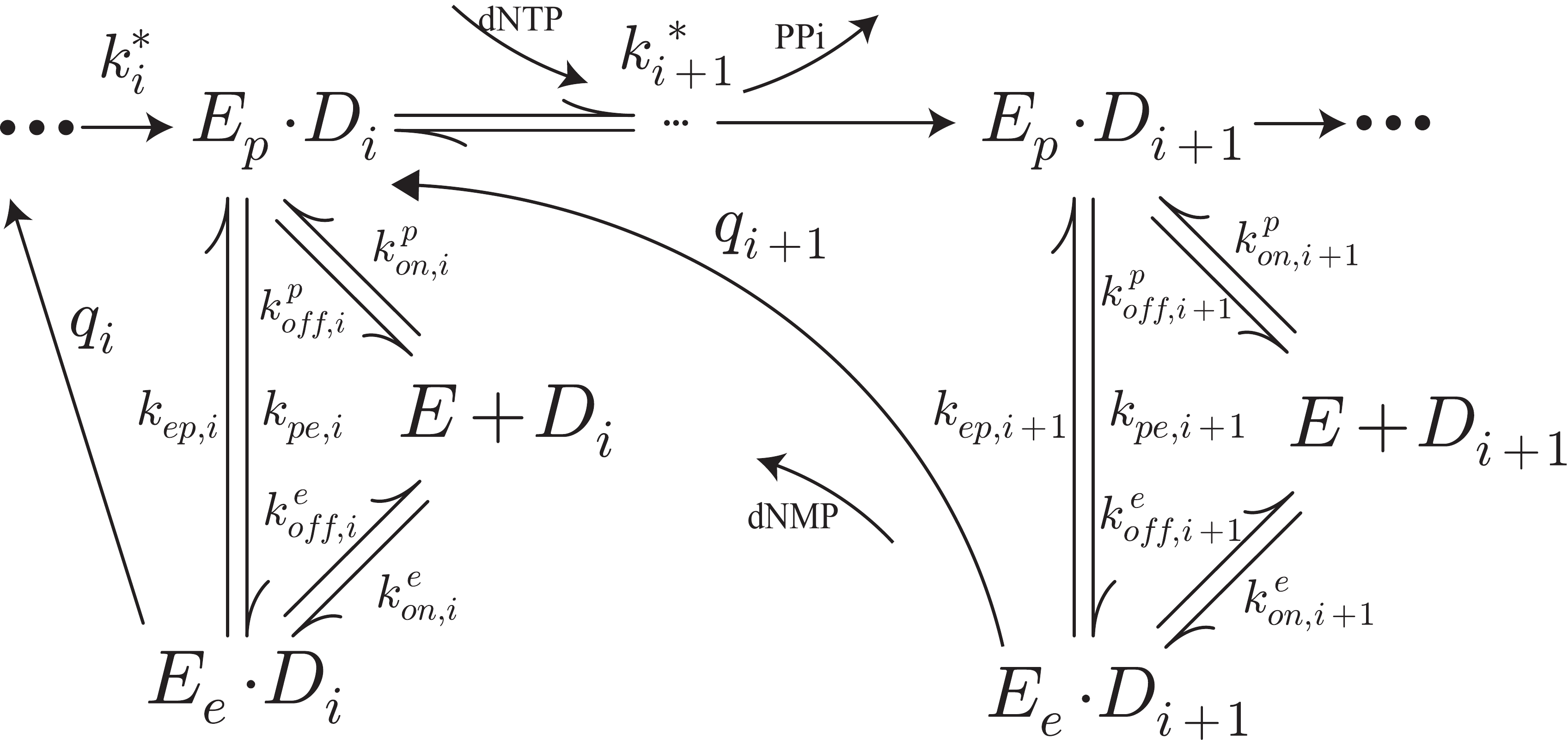}
\caption{The multi-step reaction scheme of $exo^{+}$-DNAP.}
\label{Fig_Exo_nt_FP}
\end{figure}

It is widely conjectured that the total fidelity of $exo^+$-DNAP consists of the initial discrimination $f_{ini}$ and the proofreading efficiency $f_{pro}$.  The former $f_{ini}$ can be well characterized by the methods introduced in the preceding sections. The latter $f_{pro}$, however, is assumed equal to $1+ k_{ex,W}/ k_{el,W}$, where $k_{ex}$ and $k_{el}$ are not well defined and may have different meanings in different assays.  Below we discuss some usual ways to characterize these rates.
The reaction scheme under discussion is shown in Fig.\ref{Fig_Exo_nt_FP}.

In this reaction scheme, before the excision, the primer terminal can transfer between \textit{Pol} and \textit{Exo} in two different ways, \textit{i.e.} the intramolecular transfer without DNAP dissociation (the transfer rates are denoted as $k_{pe}$ and $k_{ep}$), and the intermolecular transfer in which DNAP can dissociate from and rebind to either \textit{Pol} or \textit{Exo} (the rates are denoted as $k_{on}$ and $k_{off}$). These two modes have been revealed by single-turnover experiments\cite{Donlin1991} and directly observed by smFRET\cite{Lamichhane2013}. $k^{*}$ is the effective incorporation rate, as explained below.

\subsubsection{Effective rates uniquely determined by the FP method}\label{Sec-exo-FP}
Applying the FP method to the kinetic equations for the reaction scheme Fig.\ref{Fig_Exo_nt_FP}, one can reduce this complex scheme to the simplified scheme Fig.\ref{Fig_FP_single_step}, with rigorously defined effective rates given below. The logic of the reduction is the same as that in  \uppercase{Results and discussion} Sec.1.2.  Details can be found in SM Sec.\uppercase\expandafter{\romannumeral 1} C.
\begin{eqnarray}\label{Exo-nt-FP-erate}
\overline{k}=k^{*}&,&
\overline{r}=\frac{\widetilde{k}_{pe}q}{\widetilde{k}_{ep}+q}\nonumber\\
\widetilde{k}_{pe}=k_{pe}+k_{p\rightarrow e}&,&
\widetilde{k}_{ep}=k_{ep}+k_{e\rightarrow p}\nonumber \\
k_{p\rightarrow e}=\frac{k^{p}_{off}k^{e}_{on}}{k^{p}_{on}+k^{e}_{on}}&,&
k_{e\rightarrow p}=\frac{k^{e}_{off}k^{p}_{on}}{k^{p}_{on}+k^{e}_{on}}
\end{eqnarray}

The rate constants can be written more explicitly such as $\overline{k}_{RR}=k^*_{RR}$, if the states of the base pairs at site $i,i-1, etc.$ are explicitly indicated. All the rate constants in the same formula have the same state-subscript. $k^{*}$ is defined by Eq.(\ref{Pol-nt-FP-ieq1}).  $k_{p\rightarrow e}$ and  $k_{e\rightarrow p}$ define the effective intermolecular transfer rates between \textit{Pol} and \textit{Exo}.
So $\widetilde{k}_{pe}$ and $\widetilde{k}_{ep}$ represent the total transfer rates \textit{via} both intramolecular and intermolecular ways.
With these effective rates, the real initial discrimination and the real proofreading efficiency can be calculated by $f_{true,ini}=k^*_{RR}/k^*_{RW}$ and $f_{true,pro}=1+\overline{r}_{RW}/k^*_{WR}$ respectively.

$f_{true,pro}$ differs much from that given by K.Johnson \textit{et al.} who may be the first to discuss the contribution of the two transfer pathways to the proofreading efficiency of T7 DNAP. Without a rigorous theoretical foundation, they gave intuitively $f_{pro}=1+(k_{pe}+\theta k^{p}_{off})_{W}/k_{el,W}$ \cite{Donlin1991}.
The effective elongation rate $k_{el,W}$ was interpreted as the steady-state incorporation velocity $v^{pol}_{s\cdot s,WR}$ (at certain \textmd{[dNTP]}), which is incorrect as will be explained in the section below.
The ambiguous quantity $\theta$ was supposed between 0 and 1 (depending on the fate of the DNA after dissociation) and, unlike $\overline{r}$, can not be expressed explicitly in terms of $k^p_{on},k^p_{off},k^e_{on},k^e_{off}$,etc. So $f_{pro}$ is not equivalent to $f_{true,pro}$.

\subsubsection{The steady-state assay can not measure the proofreading efficiency}\label{Sec-exo-ss}
Because of the co-existence of the polymerase activity and the exonuclease activity, reaction schemes consisting of a single-nucleotide incorporation and a single-nucleotide excision are theoretically unacceptable and also impossible to implement in experiments. So the usual steady-state assay does not apply to $exo^+$-DNAP. It's also improper to define the elongation probability by imposing steady-state assumptions to such unrealistic reaction models, as given by Eq.(6) in Ref.\cite{Fersht1979}.
Nevertheless, the steady-state assay can still be employed to study the polymerase and exonuclease separately.

\begin{figure}[H]
\centering
\includegraphics[width=7.3cm]{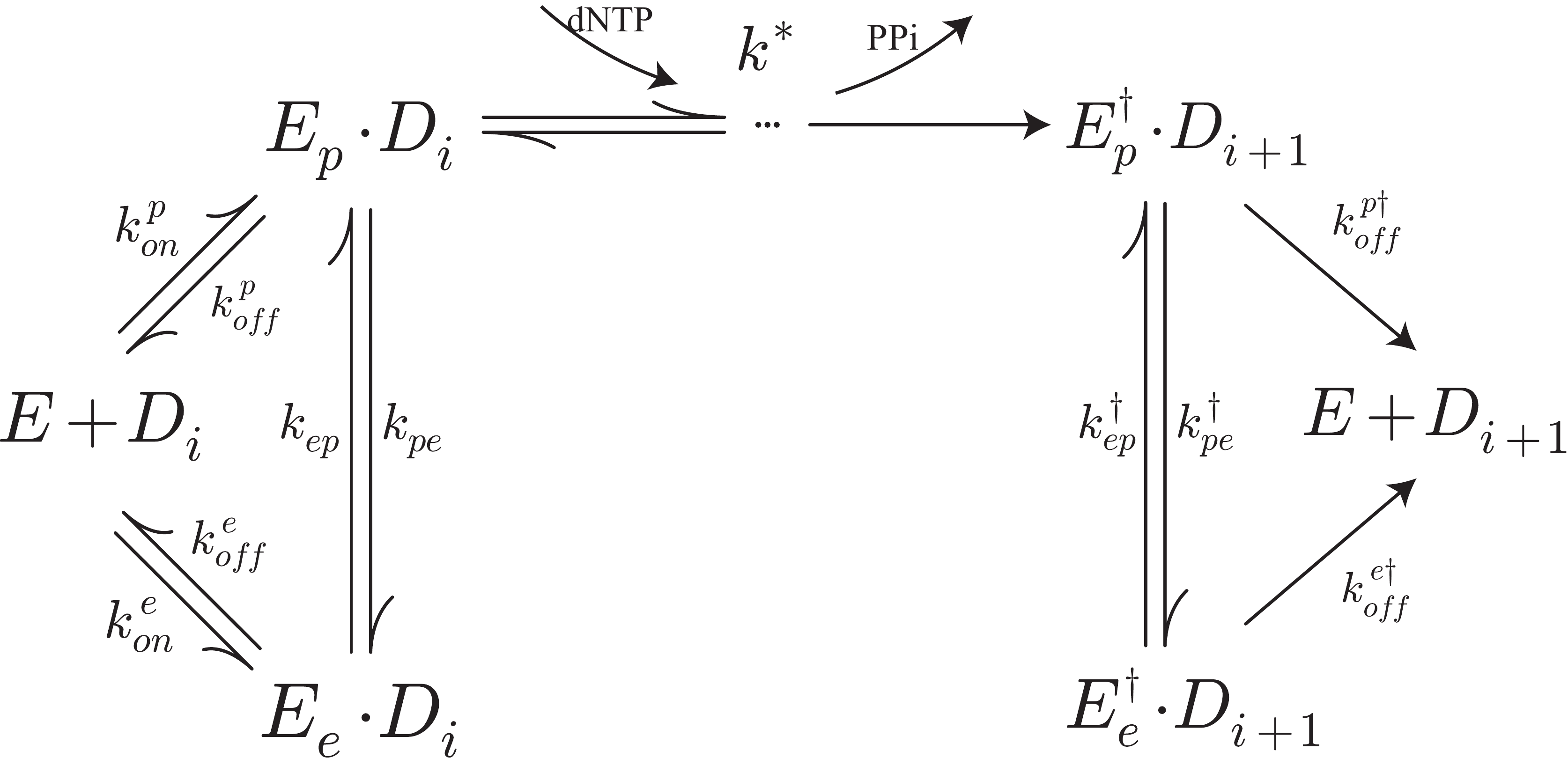}
\caption{The reaction scheme for the steady-state assay to measure the specificity constant of the nucleotide incorporation reaction of DNAP with deficient exonuclease domain.}
\label{Fig_Exo_nt_ss_pol}
\end{figure}

When mixed with the $exo^+$-DNAP, the substrate DNA can bind either to the polymerase domain or to the exonuclease domain. For some $exo^-$-DNAP, the exonuclease domain may exist and still be able to bind (but not excise) the substrate DNA, which is not discussed in preceding sections. For the steady-state assays of dNTP incorporation by such DNAPs, the reaction scheme becomes complicated (Fig.\ref{Fig_Exo_nt_ss_pol}. Again, the last enzyme dissociation steps are reasonably assumed irreversible under steady-state condition).

Under the steady-state condition, one can easily compute the specific constant
\begin{eqnarray}\label{Exo-nt-ss-specificity}
\frac{k_{cat}\textmd{[dNTP]}}{K_m}=\frac{k^*}{K^{'}_{s\cdot s}}
\end{eqnarray}
$K^{'}_{s\cdot s}=1+k_{pe}/k_{ep}+k^{p}_{off}/k^{p}_{on}$,
$k^p_{on}=k^{p0}_{on}\textmd{[DNA]}$. So the fidelity is given by
\begin{eqnarray}\label{Exo-nt-ss-fidelity-pol}
f_{ini}\equiv\frac{(k_{cat}\textmd{[dNTP]}/K_m)_{R_{i-1}R_{i}}}{(k_{cat}\textmd{[dNTP]}/K_m)_{R_{i-1}W_{i}}}
=\frac{k^*_{R_{i-1}R_{i}}}{k^*_{R_{i-1}W_{i}}}
\end{eqnarray}
which is exactly $f_{true,ini}$. Combined with Eq.(\ref{Pol-nt-ss-specificity2}), we conclude that the form of Eq.(\ref{Exo-nt-ss-specificity}) is universal: $k^*$ represents the effective incorporation rate of the subprocess beginning from dNTP binding (DNAP dissociation is not involved), and $K^{'}_{s\cdot s}$ is a simple function of the equilibrium constants of all steps before dNTP binding, no matter how complex the reaction scheme is.  Since $K^{'}_{s\cdot s}$ depends only on the identity and the state of the primer terminal but not on the next incoming dNTP, the enzyme specificity is indeed equal to $f_{true,ini}$ in general.

There were also some studies trying the steady-state assay to define the effective elongation rate $\overline{k}_{WR}$ and the effective excision rate $\overline{r}_{RW}$.  For instance, some works used the specific constant\cite{Donlin1991,Wong1991} or the maximal turn-over rate $k_{cat}$ \cite{Wingert2013} as $\overline{k}_{WR}$.  As shown above, however, $\overline{k}$ is not equal to the specific constant or $k_{cat}$.  So the steady-state assay fails in principle to measure $\overline{k}_{WR}$, unless $K^{'}_{s\cdot s}\approx1$. This condition is met by T7 DNAP ($k_{pe}\ll k_{ep}$ and $k^p_{off}\ll k^p_{on}$ were observed for the mismatched terminal\cite{Donlin1991}), and may even be generally met since replicative DNAPs are believed to be highly processive (\textit{i.e.} $k^p_{off}\ll k^p_{on}$ ) and always tend to bind DNA preferentially at the polymerase domain (\textit{i.e.} $k_{pe} / k_{ep}\ll 1$). So the the specific constant, but not $k_{cat}$, might be used in practice as $\overline{k}_{WR}$.

\begin{figure}[H]
\centering
\includegraphics[width=5.8cm]{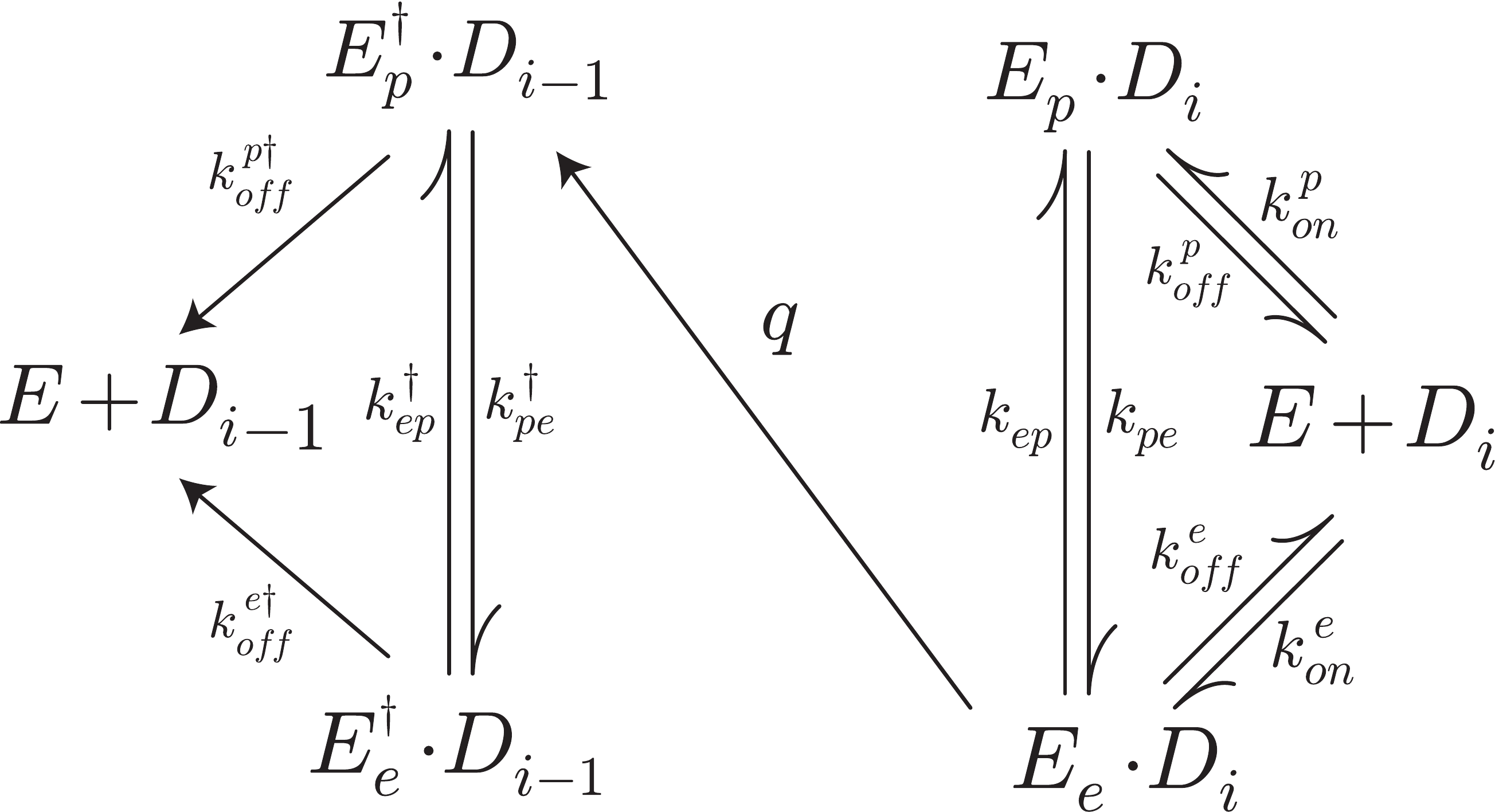}
\caption{The reaction scheme for the steady-state assay to measure the effective excision rate of $exo^+$-DNAP.}
\label{Fig_Exo_nt_ss_exo}
\end{figure}

In the steady-state assay of the excision reaction (Fig.\ref{Fig_Exo_nt_ss_exo}), one may measure the initial velocity $v^{exo}_{s\cdot s}$ and interpret it as the effective excision rate $\overline{r}$ \cite{Wingert2013}. Whereas  $v^{exo}_{s\cdot s}$ is determined by all the rate constants in Fig.\ref{Fig_Exo_nt_ss_exo},  some rate constants like $k^{p\dag}_{off}$ and $k^{\dag}_{pe}$ are absent from $\overline{r}$. So in principle $v^{exo}_{s\cdot s}$  is not equal to $\overline{r}$.  Under some conditions, $e.g.$ $k^e_{on}\ll k^p_{on}$ and the dissociation of the enzyme from the substrate is fast enough after the excision, the initial velocity $v^{exo}_{s\cdot s}$ may be approximately equal to $\overline{r}$ (details can be found in SM Sec.\uppercase\expandafter{\romannumeral 2} C). But these conditions may not be met by real DNAP,  $e.g.$, $k^e_{on}>k^p_{on}$ was observed for T7 polymerase\cite{Donlin1991}.  Unless there are carefully designed control tests to provide compelling evidence,  $v^{exo}_{s\cdot s}$ itself is not a good measure for $\overline{r}$.

One can also change the concentration of the substrate DNA to obtain the specific constant of the excision reaction in experiments\cite{Vashishtha2015}, as can be shown theoretically
\begin{eqnarray}\label{Exo-nt-ss-wrong_specificity}
\frac{k^{''}_{cat}\textmd{[DNA]}}{K^{''}_{m}}=\frac{q(k_{pe}k^{p}_{on}/(k_{pe}+k^{p}_{off})+k^{e}_{on})}{q+k_{ep}k^{p}_{off}/(k_{pe}+k^{p}_{off})+k^{e}_{off}}
\end{eqnarray}
which is just irrelevant to $\overline{r}_{RW}$.  It can be shown further in any case
\begin{eqnarray}\label{Exo-nt-ss-fidelity-exo2}
\frac{(k^{''}_{cat}\textmd{[DNA]}/K^{''}_{m})_{R_{i-1}W_{i}}}{(k_{cat}\textmd{[dNTP]}/K_m)_{W_{i}R_{i+1}}}
\neq\frac{\overline{r}_{R_{i-1}W_{i}}}{\overline{k}_{W_{i}R_{i+1}}}
\end{eqnarray}

Details can be found in SM Sec.\uppercase\expandafter{\romannumeral 2} C. In the experiment to estimate $f_{pro}$ for ap-polymerse\cite{Wingert2013}, the authors wrongly interpreted $\overline{k}_{WR}$ and $\overline{r}_{RW}$ as $k_{cat}$ and $v^{exo}_{s\cdot s}$ respectively , and gave that $excision/elongation=(v^{exo}_{s\cdot s})_{R_{i-1}W_{i}}/(k_{cat})_{W_{i}R_{i+1}}$. They thought this measure roughly reflects the true fidelity. Now it is clear that the two quantities are completely unrelated.

\subsubsection{The transient-state assay can measure the proofreading efficiency conditionally}\label{Sec-exo-ts}
\begin{figure}[H]
\centering
\includegraphics[width=5.2cm]{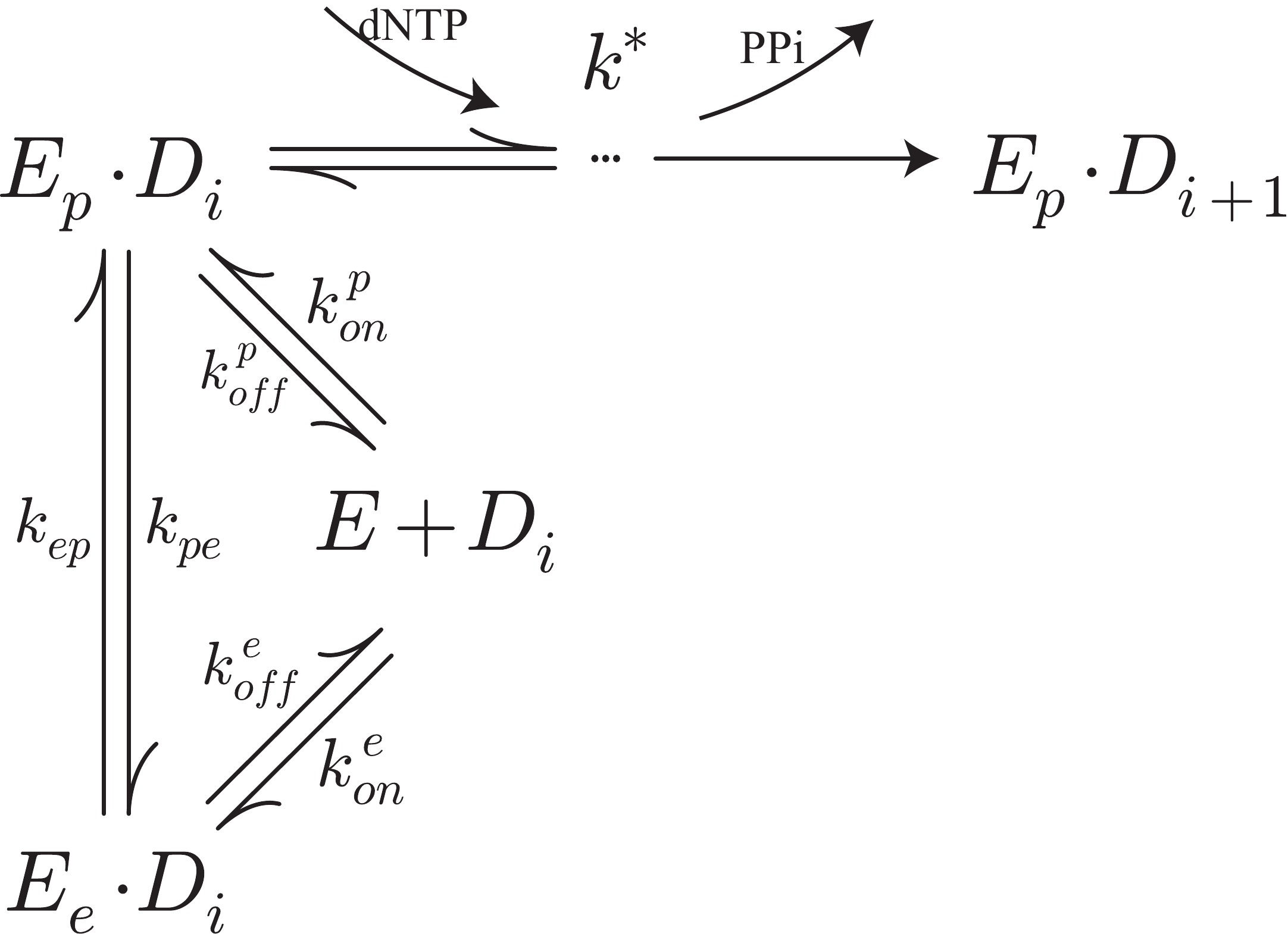}
\caption{The reaction scheme for the transient-state assay to measure the specificity constant of the nucleotide incorporation reaction of DNAP with deficient exonuclease domain.}
\label{Fig_Exo_nt_ts_pol}
\end{figure}

When DNA can bind to either the polymerase domain or the deficient exonuclease domain, the scheme for the transient-state assay of the dNTP incorporation is depicted in Fig.\ref{Fig_Exo_nt_ts_pol}.  By the same logic presented in preceding sections, the specificity constant defined by transient-state assays can be written as
\begin{eqnarray}\label{Exo-nt-ts-specificity}
\frac{k_{pol}\textmd{[dNTP]}}{K_d}=\frac{k^*}{K^{'}_{t\cdot s}}
\end{eqnarray}
$K^{'}_{t\cdot s}=1+k_{pe}/k_{ep}+k^{p}_{off}/k^{p}_{on}$.
$k^{p}_{on}=k^{p0}_{on}[E_0]$ .
This is exactly the same as Eq.(\ref{Exo-nt-ss-specificity}).
So, like the steady-state assay, the transient-state assay also applies in general to estimate $f_{true,ini}$.  Additionally, the specificity constant, but not $k_{pol}$,  can be used to estimate $\overline{k}$ when $K^{'}_{t\cdot s}\sim 1$, as mentioned in  the above section.

The transient-state assay of the exonuclease activity is often done under single-turnover conditions. The time course of product accumulation or substrate consumption is fitted by a single exponential or a double exponential to give one or two characteristic rates. In the single exponential case, the rate is simply taken as the effective excision rate $\overline{r}$. In the double exponential case, however, there is no criteria which one to select. This causes large uncertainty since the two rates often differ by one or more orders of magnitude. In the experiment of T7 polymerase\cite{Donlin1991}, two types of excision reactions were conducted, with or without preincubation of DNA and DNAP. A single characteristic rate was obtained in the former, while two rates were obtained in the latter where the smaller one almost equals to the rate in the former case. So this smaller one was selected as $\overline{r}$. In the experiment of human mitochondrial DNAP\cite{Johnson2001,Johnson2001b}, however, the larger one of the two fitted exponents was selected in some cases.  For instance, two fitted exponents of the excision reaction of the substrate 25x1/45 (DNA contains a single mismatch in the primer terminal) are $1.1s^{-1}$ and $0.04s^{-1}$ and the former was selected as $\overline{r}$\cite{Johnson2001}. Different choices of the exponents can result in estimates of $\overline{r}$ differing by orders of magnitude.
In the following, we show that the smallest of the fitted exponents may probably be equal to $\overline{r}$ under some conditions.

\begin{figure}[H]
\centering
\includegraphics[width=4cm]{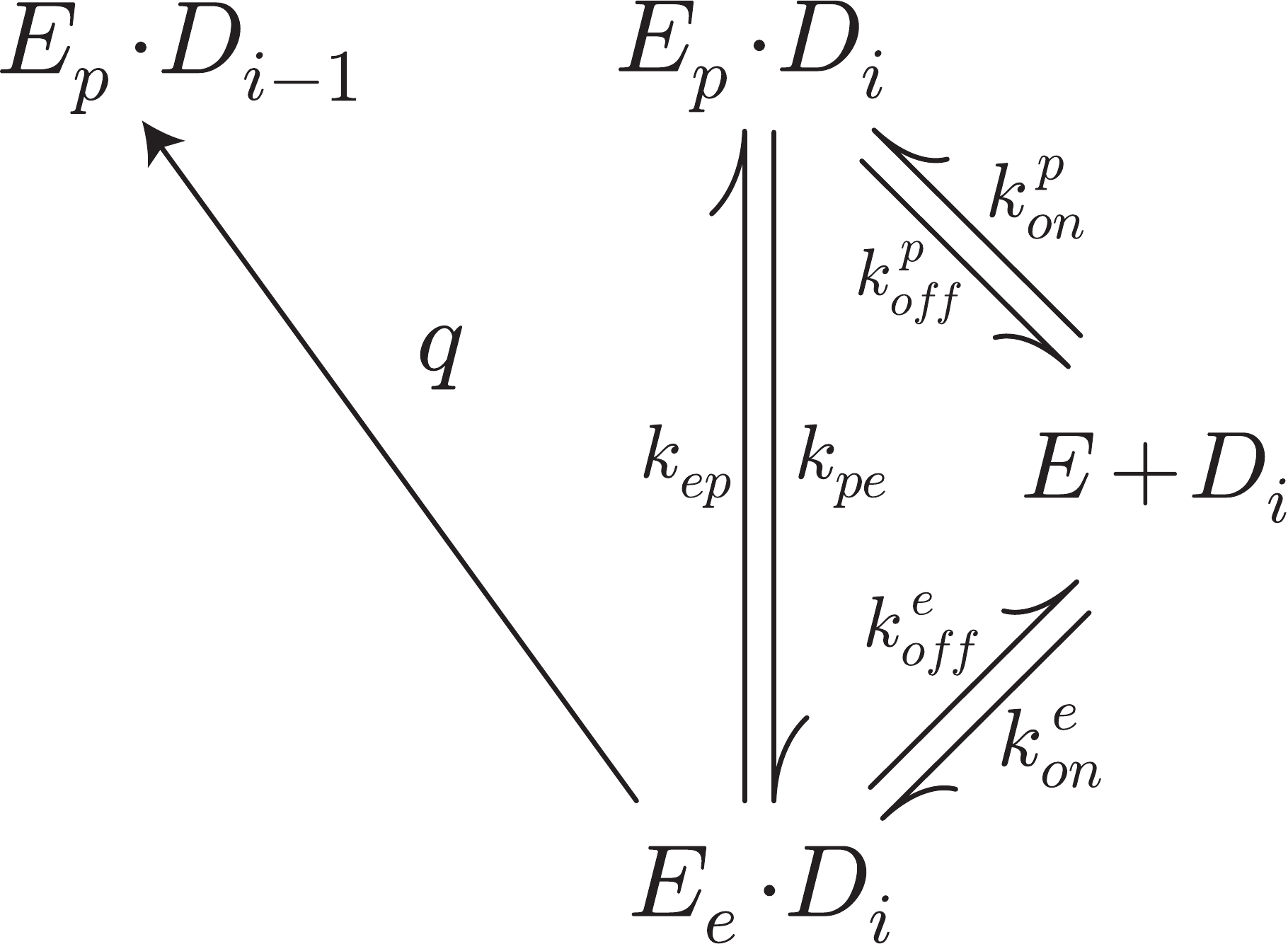}
\caption{The reaction scheme for the transient-state assay to measure the effective excision rate of $exo^+$-DNAP.}
\label{Fig_Exo_nt_ts_exo}
\end{figure}
The minimal scheme for the transient-state assay of the excision reaction is depicted in Fig.\ref{Fig_Exo_nt_ts_exo}. By solving the corresponding kinetic equations, one can get three characteristic rates and the smallest one is given by
\begin{eqnarray}\label{Exo-nt-ts-vexo1}
&&v^{exo}_{t\cdot s}=\nonumber\\&&\frac{k_{pe}q(k^{p}_{on}+k^{e}_{on})+q k^{p}_{off}k^{e}_{on}}
{(k^{p}_{on}+k^{e}_{on})(q+k_{pe}+k_{ep})+k^{e}_{off}k^{p}_{on}+k^{p}_{off}k^{e}_{on}+\epsilon}\nonumber\\
\end{eqnarray}
Here $\epsilon=k_{pe}q+k_{ep}k^{p}_{off}+k_{pe}k^{e}_{off}+q k^{p}_{off}$,  $k_{on}=k^{0}_{on}[E]$. When the concentration of DNAP is large enough to ensure $k^{p}_{on}>k^{p}_{off},k^{e}_{on}>k^{e}_{off},k^{p}_{on}>k_{pe}$ and $\epsilon \approx 0$ (compared to other terms in the denominator), Eq.\ref{Exo-nt-ts-vexo1} can be simplified as
\begin{eqnarray}\label{Exo-nt-ts-vexo2}
v^{exo}_{t\cdot s}\approx\frac{\widetilde{k}_{pe}q}{\widetilde{k}_{pe}+\widetilde{k}_{ep}+q}
\end{eqnarray}
If $\widetilde{k}_{ep}>\widetilde{k}_{pe}$, which is met if DNA binds preferentially to the polymerase domain, then we get $v^{exo}_{t\cdot s}\approx \overline{r}$ (of the same order of magnitude), $\overline{r}$ is defined in Eq.(\ref{Exo-nt-FP-erate}). So, if the real excision reaction follows the minimal scheme, $v^{exo}_{t\cdot s}$ may be interpreted as $\overline{r}$. In the experiment of human mitochondrial DNAP \cite{Johnson2001,Johnson2001b}, the author adopted this interpretation, but used $k_{pol}$ as the effective elongation rate, and calculated the proofreading efficiency as $ (v^{exo}_{t\cdot s})_{RW}/(k_{pol})_{WR}$.  It is now clear that this quantity is not a proper measure of $f_{true,pro}\big(=(v^{exo}_{t\cdot s})_{RW}/k^*_{WR}\big)$. Here $k^*_{WR}$ may probably be replaced by $(k_{pol}\textmd{[dNTP]}/K_{d})_{WR}$, if $k^p_{on} > k^p_{off}$ and $k_{ep} > k_{pe}$ .

It is worth emphasizing that the interpretation of $v^{exo}_{t\cdot s}$ is severely model-dependent. The reaction scheme could be more complicated than the minimal model in Fig.\ref{Fig_Exo_nt_ts_exo}, $e.g.$ there may be multiple substeps in the intramolecular transfer process since the two domains are far apart (2-4 nm\cite{Bebenek2018}), particularly when there are buried mismatches in the primer terminal.  For any complex scheme, one can calculate the smallest characteristic rate $v^{exo}_{t\cdot s}$ and the real excision rate $\overline{r}$. These two functions always differ greatly (examples can be found in SM Sec.\uppercase\expandafter{\romannumeral 3} D).  So the single-turnover assay \textit{per se} is not a universally reliable method to measure the effective excision rate. Is there a model-independent method to measure such excision rates?  Below we suggest a possible single-molecule assay.

\subsubsection{The single-molecule assay measures the ture fidelity}\label{Sec-exo-sm}
Similar to \uppercase{Results and discussion} Sec.1.5, a single-molecule assay can be proposed to directly measure the effective rates $\overline{k}$ and $\overline{r}$, if the states in Fig.\ref{Fig_Exo_nt_ts_pol} and Fig.\ref{Fig_Exo_nt_ts_exo} can be well defined in the experiments. For instance, the states $E_p\cdot D$, $E_e\cdot D$ and $E+D$ can be clearly resolved by smFRET \cite{Markiewicz2012,Lamichhane2013}.

To measure $\overline{k}$, the experiment is initiated by mixing DNAP and dNTP to the single molecule DNA. If $E_p\cdot D_i$, $E_e\cdot D_i$, $E+D_i$ and $E_p\cdot D_{i+1}$ in Fig.\ref{Fig_Exo_nt_ts_pol} can be distinguished in the nucleotide incorporation process, then the residence time at $E_p\cdot D_i$ can be counted from a time window of the state-switching trajectory of the enzyme-DNA complex with the starting point $E_p\cdot D_i$ and the ending point $E_p\cdot D_{i+1}$. Collecting sufficient samples to obtain the averaged residence time $\Gamma_{p,i}$, one can get $k^*_{i+1}=1/\Gamma_{p,i}$ or $k^{*0}_{i+1}=1/(\Gamma_{p,i}[d\alpha_{i+1}TP])$ by the FP analysis

The measurement of $\overline{r}$ follows the same logic.  The experiment is initiated by adding DNAPs to the single molecule DNA.  If the states $E_p\cdot D_i$, $E_e\cdot D_i$, $E+D_i$ and $E_p\cdot D_{i-1}$ in Fig.\ref{Fig_Exo_nt_ts_exo} can be distinguished in the excision process, the state-switching trajectory between the starting point $E_p\cdot D_i$ and the ending point $E_p\cdot D_{i-1}$ can be recorded. Then the averaged residence time $\Gamma_{p,i}$ at $E_p\cdot D_i$ is obtained, which gives $\overline{r}_{i}=1/\Gamma_{p,i}$. Sometimes, however, the excision may occur without visiting $E_p\cdot D$.  The trajectory recorded in such cases are not taken for the averaging. Detailed explanations can be found in SM Sec.\uppercase\expandafter{\romannumeral 4} B.  This analysis also applies to more complex reaction mechanisms and one can always get $\overline{r}_{i}=1/\Gamma_{p,i}$.

\subsection{More realistic models including DNAP translocation}\label{translocatioin}
So far we have not considered the important step, DNAP translocation, in the above kinetic models. Goodman \textit{et al.} had discussed the effect of translocation on the transient-state gel assay very early\cite{Goodman1993}, and recently DNAP translocation has been observed for phi29 DNAP by using nanopore techniques\cite{Dahl2012,Lieberman2012,Lieberman2013,Lieberman2014} or optical tweezers\cite{Morin2015}.  However, so far there is no any theory or experiment to study the effect of translocation on the replication fidelity.

\begin{figure}[H]
\centering
\includegraphics[width=6cm]{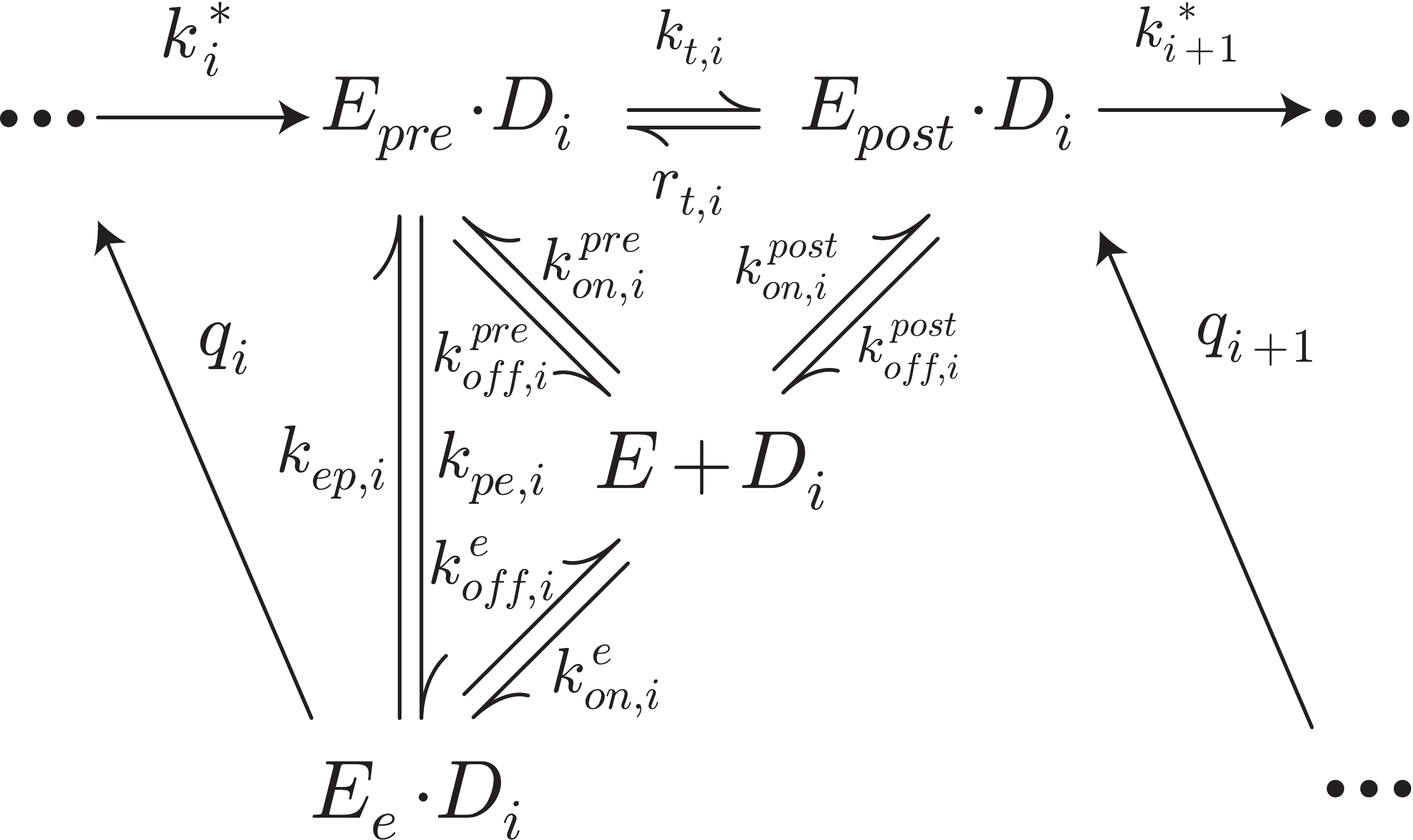}
\caption{The multi-step reaction scheme of $exo^+$-DNAP including the translocation step.}
\label{Fig_Exo_t_FP}
\end{figure}

By using optical tweezers, Morin \textit{et al.} had shown that DNAP translocation is not powered by PPi release or dNTP binding\cite{Morin2015} and it's indeed a thermal ratchet process. So the minimal scheme accounting for DNAP translocation can be depicted as Fig.\ref{Fig_Exo_t_FP}. $k_t$ and $r_t$ are the forward and the backward translocation rate respectively. $E_{pre}\cdot D_{i}$ and $E_{post}\cdot D_{i}$ indicate the pre-translocation and the post-translocation state of DNAP respectively. Here, the primer terminal can only switch intramolecularly between $E_e$ and $E_{pre}$ (but not $E_{post}$), according to the experimental observation \cite{Lieberman2014}.
We also assume DNAP can bind DNA either in state $E_{pre}\cdot D_{i}$ or in state $E_{post}\cdot D_{i}$ with possibly different binding rates and dissociation rates.

This complex scheme can be reduced to the simplified scheme  Fig.\ref{Fig_FP_single_step} by using the FP analysis. The obtained effective rates
are briefly written as $\overline{k}=k^*(1-q\eta^{'}\big/\xi)$ and $\overline{r}=q\eta\big/\xi$. Here $\eta,\eta^{'},\xi$ are complex functions of all the rate constants in Fig.\ref{Fig_Exo_t_FP} except $k^*$, which are too complex to be given here (see details in SM Sec.\uppercase\expandafter{\romannumeral 5} A).  These effective rates are much different from that defined by steady-state or transient-state assays.
Below we only show the difference between the calculated $f_{true}$ and the operationally-defined $f_{t\cdot s}$ in transient-state assays.

\begin{eqnarray}\label{Exo-t-fidelity}
(f_{true})_{i}&=&(f_{true,ini})_{i}(f_{true,pro})_{i}\nonumber\\
(f_{t\cdot s})_{i}&=&(f_{t\cdot s,ini})_{i}(f_{t\cdot s,pro})_{i}
\end{eqnarray}

The initial discrimination can be precisely measured by transient-state assays, $(f_{true,ini})_{i}=(f_{t\cdot s,ini})_{i}=k^*_{R_{i-1}R_{i}}/k^*_{R_{i-1}W_{i}}$. But the two proofreading efficiencies are hugely different, $(f_{t\cdot s,pro})_{i}\neq(f_{true,pro})_{i}$. The complex functions $(f_{true,pro})_{i}$ and $(f_{t\cdot s,pro})_{i}$ are given in SM Sec.\uppercase\expandafter{\romannumeral 5} A,B. They may be approximately equal only under some extreme conditions, \textit{e.g.}
$k_t\gg k^{pre}_{off}$, $k_t\gg k_{pe}$ and $r_t>k^{post}_{off}$.  This may be true when the terminal is in the matched state, so the translocation is in fast equilibrium and the states $pre$ and $post$ can be treated as a single state, as always assumed in conventional gel assays or other ensemble assays. But these conditions may not be met if there is a terminal mismatch or a buried mismatch which may slow down the translocation \cite{Ren2016}. In such cases, $f_{t\cdot s}$ is quite different from $f_{ture}$.  To reliably estimate $f_{ture}$, we suggest the following single-molecule assay to directly measure the effective rates.

First, if the states \textit{pre}  and \textit{post} cannot be distinguished in the experiment, indicating that the translocation is a fast process, the assays presented in preceding sections ( \uppercase{Results and discussion} Sec.1.5 or Sec.2.4) can be used.

\begin{figure}[H]
\centering
\includegraphics[width=6.3cm]{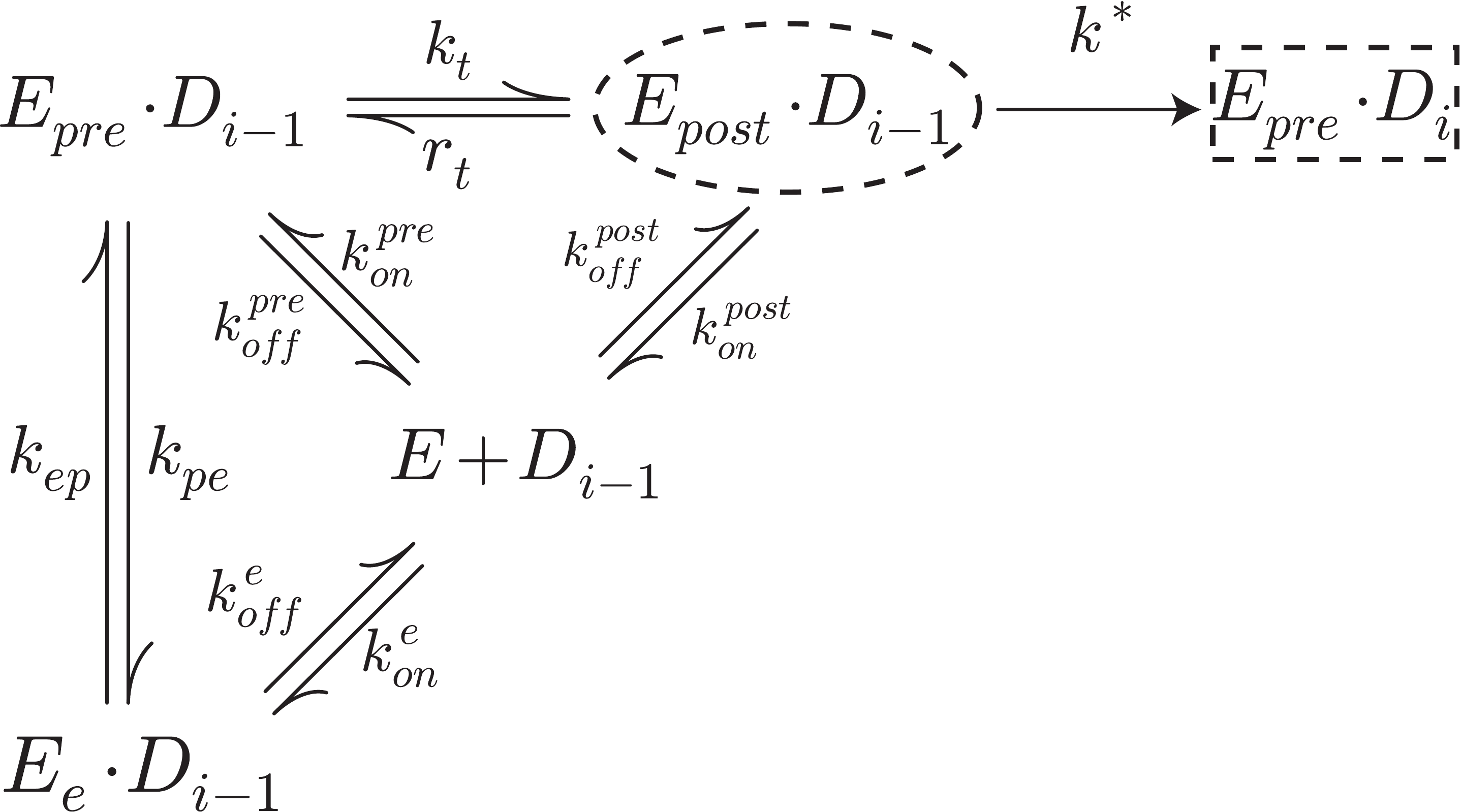}
\caption{The reaction scheme for the suggested single-molecule experiment to measure $k^*$. The dashed circle represents the starting point, and the dashed rectangle represents the ending point.}
\label{Fig_Exo_t_sm_pol}
\end{figure}

Second, if the translocation is a relatively slow process, either \textit{pre}  or \textit{post} can be directly observed (\textit{e.g.} for Dpo4 polymerase by smFRET \cite{Brenlla2014}), then the effective incorporation rate $\overline{k}$ is no longer $k^*$ but $k^*(1-q\eta^{'}\big/\xi)$.  However, it's hard to obtain this effective rate in a single measurement, since it consists of both the polymerase and the exonuclease contributions. Fortunately we can measure the factors $k^*$ and $1-q\eta^{'}\big/\xi$ separately.  The measurement of $k^*$ is basically the same as that given in  \uppercase{Results and discussion} Sec.1.5 and Sec.2.4. The reaction scheme is shown in Fig.\ref{Fig_Exo_t_sm_pol}. The experiment is initiated by mixing DNAP and dNTP to the single molecule DNA. The time trajectory between the starting point $E_{post}\cdot D_{i-1}$ and the ending point $E_{post}\cdot D_{i}$ is selected, if $E_{post}\cdot D_{i-1}$, $E_{post}\cdot D_{i}$ and other states can be well distinguished. Then the average residence time at $E_{post}\cdot D_{i-1}$ gives $k^*_{i}=1/\Gamma_{post,i-1}$ or $k^{*0}_{i}=1/(\Gamma_{post,i-1}[d\alpha_{i}TP])$, which defines $f_{true,ini}$.
Detailed explanations can be found in SM Sec.\uppercase\expandafter{\romannumeral 5} C.

\begin{figure}[H]
\centering
\includegraphics[width=6cm]{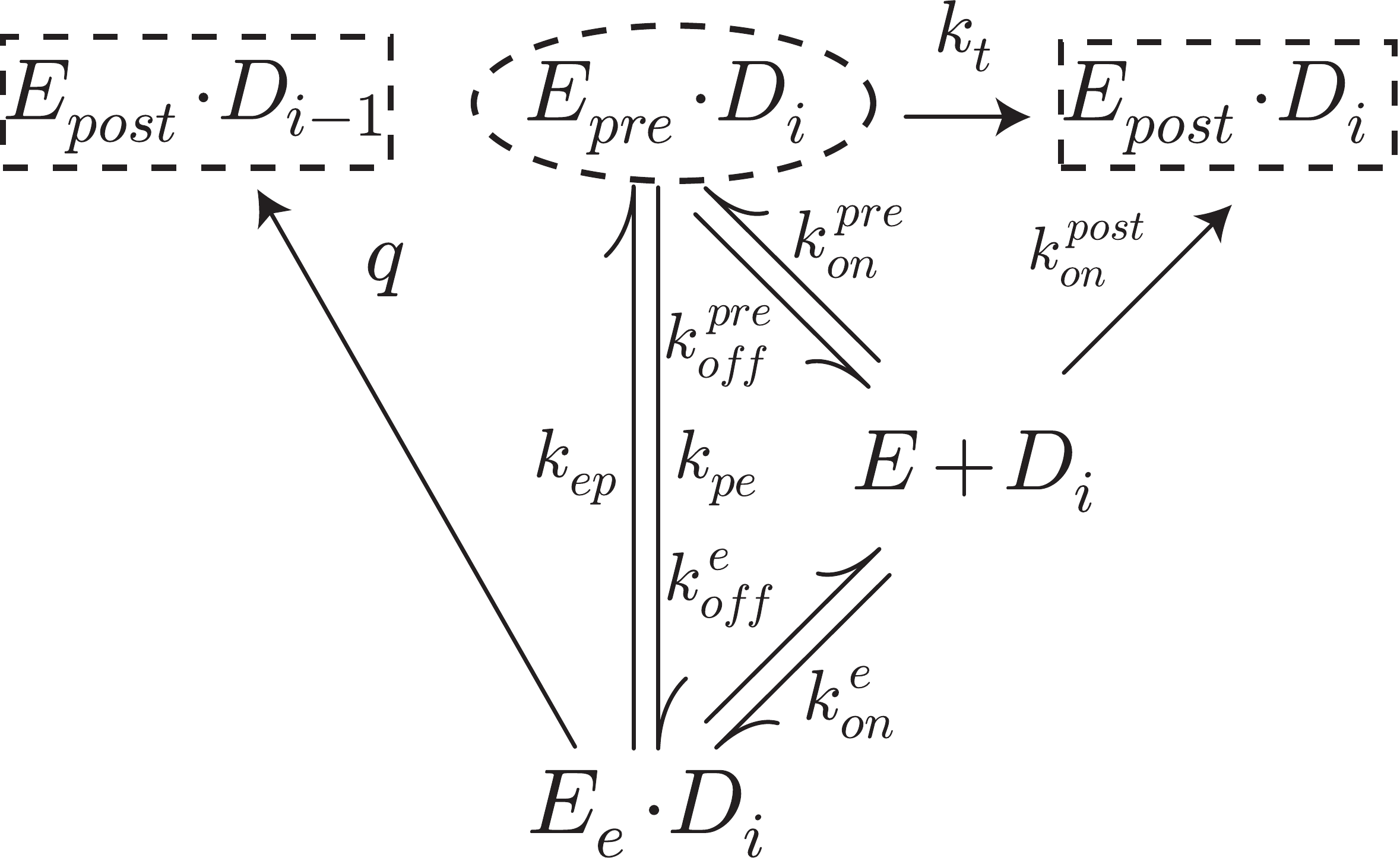}
\caption{The suggested reaction scheme to interpret the factor $q\eta^{'}\big/\xi$, based on FP analysis. The dashed circle represents the starting point, and the dashed rectangle represents the ending point.}
\label{Fig_Exo_t_sm_pol2}
\end{figure}

The logic to measure $q\eta^{'}\big/\xi$ is given below, as shown in Fig.\ref{Fig_Exo_t_sm_pol2}.

1. The experiment is initiated by mixing DNAP with DNA.

2. Record the state-switching trajectory of the complex. It may go directly to $E_{post}\cdot D_{i-1}$ without visiting $E_{pre}\cdot D_{i}$. Or it may arrive at $E_{pre}\cdot D_{i}$ \textit{via} whatever pathway before the excision, and then go to $E_{post}\cdot D_{i-1}$ (with or without visiting $E_{post}\cdot D_{i}$).  We collect trajectories of the latter case, and denote $E_{pre}\cdot D_{i}$ as the starting point (it may be visited repeatedly), $E_{post}\cdot D_{i}$ and $E_{post}\cdot D_{i-1}$ as the two ending points.

3. Select all the windows from the trajectories, which are between the starting point and either ending point. The windows are classified in two types, \textit{i.e.} between $E_{pre}\cdot D_{i}$ and $E_{post}\cdot D_{i}$, or  between $E_{pre}\cdot D_{i}$ and $E_{post}\cdot D_{i-1}$ without visiting $E_{post}\cdot D_{i}$.

4. Count the total number of either type of window $n_{post,i}$, $n_{post,i-1}$, and one gets $ n_{post,i-1}\Big/ \Big( n_{post,i-1}+n_{post,i} \Big)= (q\eta^{'}\big/\xi)_i$. Details can be found in SM Sec.\uppercase\expandafter{\romannumeral 5} C.

\begin{figure}[H]
\centering
\includegraphics[width=6.0cm]{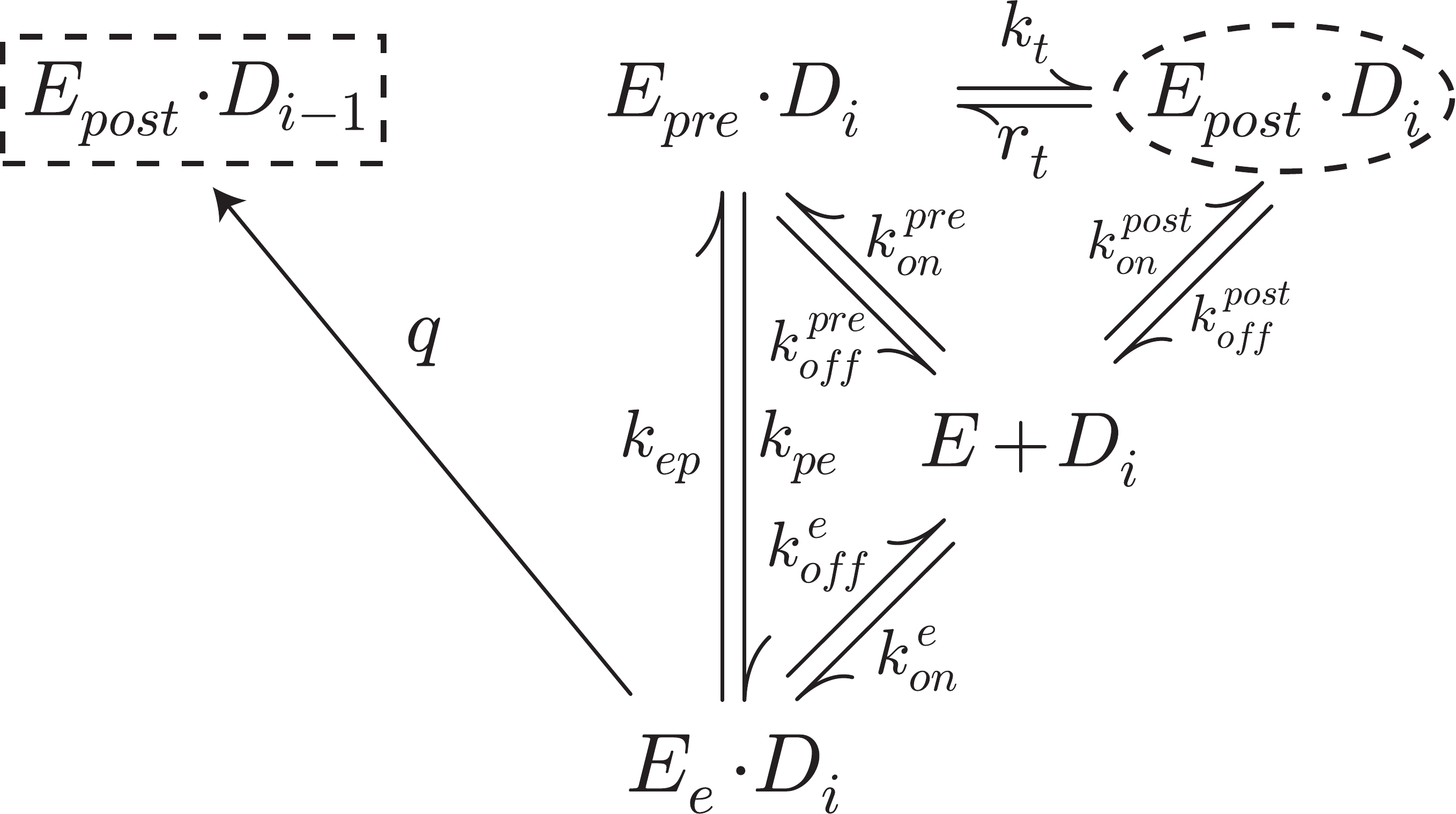}
\caption{The reaction scheme for the suggested single-molecule assay to measure $\overline{r}$. The dashed circle represents the starting point, and the dashed rectangle represents the ending point.}
\label{Fig_Exo_t_sm_exo}
\end{figure}

The measurement of $\overline{r}$ follows the same logic in  \uppercase{Results and discussion} Sec.2.4. The reaction scheme is shown in Fig.\ref{Fig_Exo_t_sm_exo}.  The experiment is initiated by adding DNAPs to the single molecule DNA. The time window selected from the trajectory is between the starting point $E_{post}\cdot D_{i}$ and the ending point $E_{post}\cdot D_{i-1}$, if $E_{post}\cdot D_{i}$, $E_{post}\cdot D_{i-1}$ and other states can be well distinguished. Then the average residence time at $E_{post}\cdot D_{i}$ gives $\overline{r}_{i}=1/\Gamma_{post,i}$. Similarly, the trajectory recorded without visiting $E_{post}\cdot D_i$ are not taken for the averaging. Detailed explanations can be found in SM Sec.\uppercase\expandafter{\romannumeral 5} C.

\section*{Summary}\label{Summary}
The conventional kinetic assays of DNAP fidelity, \textit{i.e.} the steady-state assay or the transient-state assay, have indicated that the initial discrimination $f_{ini}$ is about $10^{4\sim 5}$ and the proofreading efficiency $f_{pro}$ is about $10^{2\sim 3}$ \cite{Bebenek2018}. Although these assays have been widely used for decades and these estimates of $f_{ini}$ and $f_{pro}$ have been widely cited in the literatures, they are not unquestionable since the logic underlying these methods are not well founded. No rigorous theories about the true fidelity $f_{ture}$ have ever been proposed, and its relation to the operationally defined $f_{s\cdot s}$ or $f_{t\cdot s}$ has never been clarified.

In this paper, we examined carefully the relations among $f_{s\cdot s}$, $f_{t\cdot s}$ and $f_{true}$, based on the the FP method recently proposed by us to investigate the true fidelity of $exo^{-}$-DNAP or $exo^{+}$-DNAP.  We conclude that these three definitions are equivalent in general for $exo^{-}$-DNAP, \textit{i.e.} either the steady-state assay or the transient-state assay can give $f_{ture}$ precisely just by measuring the specificity constant ($k_{cat}/K_m$ or $k_{pol}/K_d$).

For $exo^{+}$-DNAP, however, the situation is more complicated. The steady-state assay or the transient-state assay can still be applied to measure the initial discrimination $f_{ini}$, as done for $exo^{-}$-DNAP (so the above cited estimates of $f_{ini}$ are reliable). But either method fails to measure the effective elongation rate and the effective excision rate and thus in principle can not characterize $f_{pro}$. So the widely cited estimates $f_{pro} \sim 10^{2\sim 3}$ are very suspicious. Our analysis shows that only if the involved rate constants meet some special conditions, the two assays can give approximately the effective incorporation rates, but only the transient-state assay can give approximately the effective excision rate.
If there are no other supporting evidences to ensure the required conditions are met, the conventional fidelity assays of $f_{pro}$ \textit{per se} are not reliable.  Of course, the transient-state method can be used to measure the rate constants of each step of the excision reaction\cite{Donlin1991} and then $f_{pro}$ can be calculated, but this is definitely a hard work.

So we proposed an alternative method, the single-molecule assay, to obtain the fidelity of either $exo^{-}$-DNAP or $exo^{+}$-DNAP by directly measure all the required effective rates, without dissecting the details of the reaction scheme.  It is hopefully a general and reliable method for fidelity assay if some key states of the enzyme-substrate complex can be well resolved by the single-molecule techniques.
In  \uppercase{Results and discussion} Sec.1.5, Sec.2.4 and Sec.3, we have designed several protocols to conduct the single-molecule experiment and data analysis, which are feasible in principle though it may be hard to implement in practice.

Last but not least, we have focused on the first-order (nearest) neighbor effects in this paper, but higher-order neighbor effects may also be important to the fidelity.
Here we take the second-order (next-nearest) neighbor effect as an example.
According to Eq.(\ref{Fid_eq_high-order}), the initial discrimination $f_{ini}$ is of the same form as that defined by Eq.(\ref{Fid-eq-pol}), so it can still be correctly given by the steady-state or the transient-state assays (in fact, these assays can measure $f_{ini}$ for any-order neighbor effects).
The proofreading efficiency $f_{pro}$ of $exo^+$-DNAP consists of two factors, $\overline{r}_{R_{i-2}R_{i-1}W_{i}}/\overline{k}_{R_{i-1}W_{i}R_{i+1}}$ and
$\overline{r}_{R_{i-1}W_{i}R_{i+1}}/\overline{k}_{W_{i}R_{i+1}R_{i+2}}$,
which can be regarded as the first-order and the second-order proofreading efficiency respectively. These two factors are both dependent on the stability of the primer-template duplex. For naked dsDNA duplex, numerous experiments have shown that a penultimate mismatch leads to much lower stability than a terminal mismatch \cite{SantaLucia2004}. This implies that a penultimate mismatch may more significantly disturb the base stacking of the primer-template conjunction in the polymerase domain and thus the forward translocation of DNAP will be slower and the $Pol$-to-$Exo$ transfer of the primer terminal will be faster, if compared with the terminal mismatch. In such cases, the second-order factor may be larger than the first-order factor.
This enhancement has been mentioned in Ref.\cite{Johnson2001}, though the steady-state and transient-state assays used in that work and thus the obtained estimates of the two factors are all questionable (as pointed out in  \uppercase{Results and discussion} Sec.2).
The single-molecule assay can be directly adopted to demonstrate the second-order effects by measuring the effective rates $\overline{r}_{R_{i-1}W_{i}R_{i+1}}, \overline{k}_{W_{i}R_{i+1}R_{i+2}}$ by the same protocols.  We hope the analysis and the suggestions presented in this paper will urge a serious examination of the conventional fidelity assays and offer some new inspirations to single-molecule experimentalists to conduct more accurate fidelity analysis.

\section*{Acknowledgments}
The authors thank the financial support by National Natural Science Foundation of China (No.11675180,11774358), the CAS Strategic Priority Research Program (No.XDA17010504), Key Research Program of Frontier Sciences of CAS (No.Y7Y1472Y61),  WIUCASYJ2020004 and WIUCASQD2020009.

\bibliography{TDdnaP_RealScheme_20200712}

\begin{thebibliography}{38}
\expandafter\ifx\csname natexlab\endcsname\relax\def\natexlab#1{#1}\fi
\expandafter\ifx\csname bibnamefont\endcsname\relax
  \def\bibnamefont#1{#1}\fi
\expandafter\ifx\csname bibfnamefont\endcsname\relax
  \def\bibfnamefont#1{#1}\fi
\expandafter\ifx\csname citenamefont\endcsname\relax
  \def\citenamefont#1{#1}\fi
\expandafter\ifx\csname url\endcsname\relax
  \def\url#1{\texttt{#1}}\fi
\expandafter\ifx\csname urlprefix\endcsname\relax\def\urlprefix{URL }\fi
\providecommand{\bibinfo}[2]{#2}
\providecommand{\eprint}[2][]{\url{#2}}

\bibitem[{\citenamefont{Trautner et~al.}(1962)\citenamefont{Trautner, Swartz,
  and Kornberg}}]{Trautner1962}
\bibinfo{author}{\bibfnamefont{T.~A.} \bibnamefont{Trautner}},
  \bibinfo{author}{\bibfnamefont{M.~N.} \bibnamefont{Swartz}},
  \bibnamefont{and} \bibinfo{author}{\bibfnamefont{A.}~\bibnamefont{Kornberg}},
  \bibinfo{journal}{Proceedings of the National Academy of Sciences of the
  United States of America} \textbf{\bibinfo{volume}{48}}, \bibinfo{pages}{449}
  (\bibinfo{year}{1962}).

\bibitem[{\citenamefont{Hall and Lehman}(1968)}]{Hall1968}
\bibinfo{author}{\bibfnamefont{Z.~W.} \bibnamefont{Hall}} \bibnamefont{and}
  \bibinfo{author}{\bibfnamefont{I.~R.} \bibnamefont{Lehman}},
  \bibinfo{journal}{Journal of Molecular Biology}
  \textbf{\bibinfo{volume}{36}}, \bibinfo{pages}{321} (\bibinfo{year}{1968}).

\bibitem[{\citenamefont{Lee et~al.}(2016)\citenamefont{Lee, Lu, Chang, Loparo,
  and Xie}}]{Lee2016}
\bibinfo{author}{\bibfnamefont{D.~F.} \bibnamefont{Lee}},
  \bibinfo{author}{\bibfnamefont{J.}~\bibnamefont{Lu}},
  \bibinfo{author}{\bibfnamefont{S.}~\bibnamefont{Chang}},
  \bibinfo{author}{\bibfnamefont{J.~J.} \bibnamefont{Loparo}},
  \bibnamefont{and} \bibinfo{author}{\bibfnamefont{X.~S.} \bibnamefont{Xie}},
  \bibinfo{journal}{Nucleic Acids Research} \textbf{\bibinfo{volume}{44}},
  \bibinfo{pages}{e118} (\bibinfo{year}{2016}).

\bibitem[{\citenamefont{de~Paz et~al.}(2018)\citenamefont{de~Paz, Cybulski,
  Marblestone, Zamft, Church, Boyden, Kording, and Tyo}}]{DePaz2018}
\bibinfo{author}{\bibfnamefont{A.~M.} \bibnamefont{de~Paz}},
  \bibinfo{author}{\bibfnamefont{T.~R.} \bibnamefont{Cybulski}},
  \bibinfo{author}{\bibfnamefont{A.~H.} \bibnamefont{Marblestone}},
  \bibinfo{author}{\bibfnamefont{B.~M.} \bibnamefont{Zamft}},
  \bibinfo{author}{\bibfnamefont{G.~M.} \bibnamefont{Church}},
  \bibinfo{author}{\bibfnamefont{E.~S.} \bibnamefont{Boyden}},
  \bibinfo{author}{\bibfnamefont{K.~P.} \bibnamefont{Kording}},
  \bibnamefont{and} \bibinfo{author}{\bibfnamefont{K.~E.} \bibnamefont{Tyo}},
  \bibinfo{journal}{Nucleic Acids Research} \textbf{\bibinfo{volume}{46}},
  \bibinfo{pages}{e78} (\bibinfo{year}{2018}).

\bibitem[{\citenamefont{Clayton and Branscomb}(1979)}]{Clayton1979}
\bibinfo{author}{\bibfnamefont{K.}~\bibnamefont{Clayton}} \bibnamefont{and}
  \bibinfo{author}{\bibfnamefont{W.}~\bibnamefont{Branscomb}},
  \bibinfo{journal}{Journal of Biological Chemistry}
  \textbf{\bibinfo{volume}{254}}, \bibinfo{pages}{1902} (\bibinfo{year}{1979}).

\bibitem[{\citenamefont{Bertram et~al.}(2010)\citenamefont{Bertram, Oertell,
  Petruska, and Goodman}}]{Bertram2010}
\bibinfo{author}{\bibfnamefont{J.~G.} \bibnamefont{Bertram}},
  \bibinfo{author}{\bibfnamefont{K.}~\bibnamefont{Oertell}},
  \bibinfo{author}{\bibfnamefont{J.}~\bibnamefont{Petruska}}, \bibnamefont{and}
  \bibinfo{author}{\bibfnamefont{M.~F.} \bibnamefont{Goodman}},
  \bibinfo{journal}{Biochemistry} \textbf{\bibinfo{volume}{49}},
  \bibinfo{pages}{20} (\bibinfo{year}{2010}).

\bibitem[{\citenamefont{Fersht}(1985)}]{Fersht1985}
\bibinfo{author}{\bibfnamefont{A.~R.} \bibnamefont{Fersht}},
  \emph{\bibinfo{title}{{Enzyme Structure and Mechanism}}}
  (\bibinfo{publisher}{W.H.Freeman {\&} Co Ltd.}, \bibinfo{year}{1985}),
  \bibinfo{edition}{2nd} ed.

\bibitem[{\citenamefont{Kati et~al.}(1992)\citenamefont{Kati, Johnson, Jerva,
  and Anderson}}]{Kati1992}
\bibinfo{author}{\bibfnamefont{W.~M.} \bibnamefont{Kati}},
  \bibinfo{author}{\bibfnamefont{K.~A.} \bibnamefont{Johnson}},
  \bibinfo{author}{\bibfnamefont{L.~F.} \bibnamefont{Jerva}}, \bibnamefont{and}
  \bibinfo{author}{\bibfnamefont{K.~S.} \bibnamefont{Anderson}},
  \bibinfo{journal}{Journal of Biological Chemistry}
  \textbf{\bibinfo{volume}{267}}, \bibinfo{pages}{25988}
  (\bibinfo{year}{1992}).

\bibitem[{\citenamefont{Johnson}(1993)}]{Johnson1993}
\bibinfo{author}{\bibfnamefont{K.~A.} \bibnamefont{Johnson}},
  \bibinfo{journal}{Annual Review of Biochemistry}
  \textbf{\bibinfo{volume}{62}}, \bibinfo{pages}{685} (\bibinfo{year}{1993}).

\bibitem[{\citenamefont{Johnson}(1992)}]{Johnson1992}
\bibinfo{author}{\bibfnamefont{K.~A.} \bibnamefont{Johnson}},
  \bibinfo{journal}{The Enzymes} \textbf{\bibinfo{volume}{20}},
  \bibinfo{pages}{1} (\bibinfo{year}{1992}).

\bibitem[{\citenamefont{Fersht et~al.}(1982)\citenamefont{Fersht, Knill-Jones,
  and Tsui}}]{Fersht1982}
\bibinfo{author}{\bibfnamefont{A.~R.} \bibnamefont{Fersht}},
  \bibinfo{author}{\bibfnamefont{J.~W.} \bibnamefont{Knill-Jones}},
  \bibnamefont{and} \bibinfo{author}{\bibfnamefont{W.~C.} \bibnamefont{Tsui}},
  \bibinfo{journal}{Journal of Molecular Biology}
  \textbf{\bibinfo{volume}{156}}, \bibinfo{pages}{37} (\bibinfo{year}{1982}).

\bibitem[{\citenamefont{Fersht}(1979)}]{Fersht1979}
\bibinfo{author}{\bibfnamefont{A.~R.} \bibnamefont{Fersht}},
  \bibinfo{journal}{Proceedings of the National Academy of Sciences of the
  United States of America} \textbf{\bibinfo{volume}{76}},
  \bibinfo{pages}{4946} (\bibinfo{year}{1979}).

\bibitem[{\citenamefont{Wingert et~al.}(2013)\citenamefont{Wingert, Parrott,
  and Nelson}}]{Wingert2013}
\bibinfo{author}{\bibfnamefont{B.~M.} \bibnamefont{Wingert}},
  \bibinfo{author}{\bibfnamefont{E.~E.} \bibnamefont{Parrott}},
  \bibnamefont{and} \bibinfo{author}{\bibfnamefont{S.~W.}
  \bibnamefont{Nelson}}, \bibinfo{journal}{Biochemistry}
  \textbf{\bibinfo{volume}{52}}, \bibinfo{pages}{7723} (\bibinfo{year}{2013}).

\bibitem[{\citenamefont{Vashishtha and Kuchta}(2015)}]{Vashishtha2015}
\bibinfo{author}{\bibfnamefont{A.~K.} \bibnamefont{Vashishtha}}
  \bibnamefont{and} \bibinfo{author}{\bibfnamefont{R.~D.}
  \bibnamefont{Kuchta}}, \bibinfo{journal}{Biochemistry}
  \textbf{\bibinfo{volume}{54}}, \bibinfo{pages}{240} (\bibinfo{year}{2015}).

\bibitem[{\citenamefont{Donlin et~al.}(1991)\citenamefont{Donlin, Patel, and
  Johnson}}]{Donlin1991}
\bibinfo{author}{\bibfnamefont{M.~J.} \bibnamefont{Donlin}},
  \bibinfo{author}{\bibfnamefont{S.~S.} \bibnamefont{Patel}}, \bibnamefont{and}
  \bibinfo{author}{\bibfnamefont{K.~A.} \bibnamefont{Johnson}},
  \bibinfo{journal}{Biochemistry} \textbf{\bibinfo{volume}{30}},
  \bibinfo{pages}{538} (\bibinfo{year}{1991}).

\bibitem[{\citenamefont{Gaspard}(2017)}]{Gaspard2017}
\bibinfo{author}{\bibfnamefont{P.}~\bibnamefont{Gaspard}},
  \bibinfo{journal}{Physical Review E} \textbf{\bibinfo{volume}{96}},
  \bibinfo{pages}{1} (\bibinfo{year}{2017}).

\bibitem[{\citenamefont{Li et~al.}(2019)\citenamefont{Li, Zheng, Shu, Ou-Yang,
  and Li}}]{Li2019}
\bibinfo{author}{\bibfnamefont{Q.~S.} \bibnamefont{Li}},
  \bibinfo{author}{\bibfnamefont{P.~D.} \bibnamefont{Zheng}},
  \bibinfo{author}{\bibfnamefont{Y.~G.} \bibnamefont{Shu}},
  \bibinfo{author}{\bibfnamefont{Z.~C.} \bibnamefont{Ou-Yang}},
  \bibnamefont{and} \bibinfo{author}{\bibfnamefont{M.}~\bibnamefont{Li}},
  \bibinfo{journal}{Physical Review E} \textbf{\bibinfo{volume}{100}}
  (\bibinfo{year}{2019}), \eprint{1901.01495}.

\bibitem[{\citenamefont{Song et~al.}(2017)\citenamefont{Song, Shu, Zhou,
  Ou-Yang, and Li}}]{Song2017}
\bibinfo{author}{\bibfnamefont{Y.~S.} \bibnamefont{Song}},
  \bibinfo{author}{\bibfnamefont{Y.~G.} \bibnamefont{Shu}},
  \bibinfo{author}{\bibfnamefont{X.}~\bibnamefont{Zhou}},
  \bibinfo{author}{\bibfnamefont{Z.~C.} \bibnamefont{Ou-Yang}},
  \bibnamefont{and} \bibinfo{author}{\bibfnamefont{M.}~\bibnamefont{Li}},
  \bibinfo{journal}{Journal of Physics Condensed Matter}
  \textbf{\bibinfo{volume}{29}}, \bibinfo{pages}{25101} (\bibinfo{year}{2017}),
  \eprint{1603.02453}.

\bibitem[{\citenamefont{Wong et~al.}(1991)\citenamefont{Wong, Patel, and
  Johnson}}]{Wong1991}
\bibinfo{author}{\bibfnamefont{I.}~\bibnamefont{Wong}},
  \bibinfo{author}{\bibfnamefont{S.~S.} \bibnamefont{Patel}}, \bibnamefont{and}
  \bibinfo{author}{\bibfnamefont{K.~A.} \bibnamefont{Johnson}},
  \bibinfo{journal}{Biochemistry} \textbf{\bibinfo{volume}{30}},
  \bibinfo{pages}{526} (\bibinfo{year}{1991}).

\bibitem[{\citenamefont{Tsai and Johnson}(2006)}]{Tsai2006}
\bibinfo{author}{\bibfnamefont{Y.~C.} \bibnamefont{Tsai}} \bibnamefont{and}
  \bibinfo{author}{\bibfnamefont{K.~A.} \bibnamefont{Johnson}},
  \bibinfo{journal}{Biochemistry} \textbf{\bibinfo{volume}{45}},
  \bibinfo{pages}{9675} (\bibinfo{year}{2006}).

\bibitem[{\citenamefont{Purohit et~al.}(2003)\citenamefont{Purohit, Grindley,
  and Joyce}}]{Purohit2003}
\bibinfo{author}{\bibfnamefont{V.}~\bibnamefont{Purohit}},
  \bibinfo{author}{\bibfnamefont{N.~D.} \bibnamefont{Grindley}},
  \bibnamefont{and} \bibinfo{author}{\bibfnamefont{C.~M.} \bibnamefont{Joyce}},
  \bibinfo{journal}{Biochemistry} \textbf{\bibinfo{volume}{42}},
  \bibinfo{pages}{10200} (\bibinfo{year}{2003}).

\bibitem[{\citenamefont{Joyce et~al.}(2008)\citenamefont{Joyce, Potapova,
  DeLucia, Huang, Basu, and Grindley}}]{Joyce2008}
\bibinfo{author}{\bibfnamefont{C.~M.} \bibnamefont{Joyce}},
  \bibinfo{author}{\bibfnamefont{O.}~\bibnamefont{Potapova}},
  \bibinfo{author}{\bibfnamefont{A.~M.} \bibnamefont{DeLucia}},
  \bibinfo{author}{\bibfnamefont{X.}~\bibnamefont{Huang}},
  \bibinfo{author}{\bibfnamefont{V.~P.} \bibnamefont{Basu}}, \bibnamefont{and}
  \bibinfo{author}{\bibfnamefont{N.~D.} \bibnamefont{Grindley}},
  \bibinfo{journal}{Biochemistry} \textbf{\bibinfo{volume}{47}},
  \bibinfo{pages}{6103} (\bibinfo{year}{2008}).

\bibitem[{\citenamefont{Santoso et~al.}(2010)\citenamefont{Santoso, Joyce,
  Potapova, {Le Reste}, Hohlbein, Torella, Grindley, and
  Kapanidis}}]{Santoso2010}
\bibinfo{author}{\bibfnamefont{Y.}~\bibnamefont{Santoso}},
  \bibinfo{author}{\bibfnamefont{C.~M.} \bibnamefont{Joyce}},
  \bibinfo{author}{\bibfnamefont{O.}~\bibnamefont{Potapova}},
  \bibinfo{author}{\bibfnamefont{L.}~\bibnamefont{{Le Reste}}},
  \bibinfo{author}{\bibfnamefont{J.}~\bibnamefont{Hohlbein}},
  \bibinfo{author}{\bibfnamefont{J.~P.} \bibnamefont{Torella}},
  \bibinfo{author}{\bibfnamefont{N.~D.} \bibnamefont{Grindley}},
  \bibnamefont{and} \bibinfo{author}{\bibfnamefont{A.~N.}
  \bibnamefont{Kapanidis}}, \bibinfo{journal}{Proceedings of the National
  Academy of Sciences of the United States of America}
  \textbf{\bibinfo{volume}{107}}, \bibinfo{pages}{715} (\bibinfo{year}{2010}).

\bibitem[{\citenamefont{Hohlbein et~al.}(2013)\citenamefont{Hohlbein, Aigrain,
  Craggs, Bermek, Potapova, Shoolizadeh, Grindley, Joyce, and
  Kapanidis}}]{Hohlbein2013}
\bibinfo{author}{\bibfnamefont{J.}~\bibnamefont{Hohlbein}},
  \bibinfo{author}{\bibfnamefont{L.}~\bibnamefont{Aigrain}},
  \bibinfo{author}{\bibfnamefont{T.~D.} \bibnamefont{Craggs}},
  \bibinfo{author}{\bibfnamefont{O.}~\bibnamefont{Bermek}},
  \bibinfo{author}{\bibfnamefont{O.}~\bibnamefont{Potapova}},
  \bibinfo{author}{\bibfnamefont{P.}~\bibnamefont{Shoolizadeh}},
  \bibinfo{author}{\bibfnamefont{N.~D.} \bibnamefont{Grindley}},
  \bibinfo{author}{\bibfnamefont{C.~M.} \bibnamefont{Joyce}}, \bibnamefont{and}
  \bibinfo{author}{\bibfnamefont{A.~N.} \bibnamefont{Kapanidis}},
  \bibinfo{journal}{Nature Communications} \textbf{\bibinfo{volume}{4}}
  (\bibinfo{year}{2013}).

\bibitem[{\citenamefont{Lamichhane et~al.}(2013)\citenamefont{Lamichhane,
  Berezhna, Gill, {Van Der Schans}, and Millar}}]{Lamichhane2013}
\bibinfo{author}{\bibfnamefont{R.}~\bibnamefont{Lamichhane}},
  \bibinfo{author}{\bibfnamefont{S.~Y.} \bibnamefont{Berezhna}},
  \bibinfo{author}{\bibfnamefont{J.~P.} \bibnamefont{Gill}},
  \bibinfo{author}{\bibfnamefont{E.}~\bibnamefont{{Van Der Schans}}},
  \bibnamefont{and} \bibinfo{author}{\bibfnamefont{D.~P.}
  \bibnamefont{Millar}}, \bibinfo{journal}{Journal of the American Chemical
  Society} \textbf{\bibinfo{volume}{135}}, \bibinfo{pages}{4735}
  (\bibinfo{year}{2013}).

\bibitem[{\citenamefont{Johnson and Johnson}(2001{\natexlab{a}})}]{Johnson2001}
\bibinfo{author}{\bibfnamefont{A.~A.} \bibnamefont{Johnson}} \bibnamefont{and}
  \bibinfo{author}{\bibfnamefont{K.~A.} \bibnamefont{Johnson}},
  \bibinfo{journal}{Journal of Biological Chemistry}
  \textbf{\bibinfo{volume}{276}}, \bibinfo{pages}{38097}
  (\bibinfo{year}{2001}{\natexlab{a}}).

\bibitem[{\citenamefont{Johnson and
  Johnson}(2001{\natexlab{b}})}]{Johnson2001b}
\bibinfo{author}{\bibfnamefont{A.~A.} \bibnamefont{Johnson}} \bibnamefont{and}
  \bibinfo{author}{\bibfnamefont{K.~A.} \bibnamefont{Johnson}},
  \bibinfo{journal}{Journal of Biological Chemistry}
  \textbf{\bibinfo{volume}{276}}, \bibinfo{pages}{38090}
  (\bibinfo{year}{2001}{\natexlab{b}}).

\bibitem[{\citenamefont{B{\c{e}}benek and Ziuzia-Graczyk}(2018)}]{Bebenek2018}
\bibinfo{author}{\bibfnamefont{A.}~\bibnamefont{B{\c{e}}benek}}
  \bibnamefont{and}
  \bibinfo{author}{\bibfnamefont{I.}~\bibnamefont{Ziuzia-Graczyk}},
  \bibinfo{journal}{Current Genetics} \textbf{\bibinfo{volume}{64}},
  \bibinfo{pages}{985} (\bibinfo{year}{2018}).

\bibitem[{\citenamefont{Markiewicz et~al.}(2012)\citenamefont{Markiewicz,
  Vrtis, Rueda, and Romano}}]{Markiewicz2012}
\bibinfo{author}{\bibfnamefont{R.~P.} \bibnamefont{Markiewicz}},
  \bibinfo{author}{\bibfnamefont{K.~B.} \bibnamefont{Vrtis}},
  \bibinfo{author}{\bibfnamefont{D.}~\bibnamefont{Rueda}}, \bibnamefont{and}
  \bibinfo{author}{\bibfnamefont{L.~J.} \bibnamefont{Romano}},
  \bibinfo{journal}{Nucleic Acids Research} \textbf{\bibinfo{volume}{40}},
  \bibinfo{pages}{7975} (\bibinfo{year}{2012}).

\bibitem[{\citenamefont{Goodman et~al.}(1993)\citenamefont{Goodman, Creighton,
  Bloom, Petruska, and Kunkel}}]{Goodman1993}
\bibinfo{author}{\bibfnamefont{M.~F.} \bibnamefont{Goodman}},
  \bibinfo{author}{\bibfnamefont{S.}~\bibnamefont{Creighton}},
  \bibinfo{author}{\bibfnamefont{L.~B.} \bibnamefont{Bloom}},
  \bibinfo{author}{\bibfnamefont{J.}~\bibnamefont{Petruska}}, \bibnamefont{and}
  \bibinfo{author}{\bibfnamefont{T.~A.} \bibnamefont{Kunkel}},
  \bibinfo{journal}{Critical Reviews in Biochemistry and Molecular Biology}
  \textbf{\bibinfo{volume}{28}}, \bibinfo{pages}{83} (\bibinfo{year}{1993}).

\bibitem[{\citenamefont{Dahl et~al.}(2012)\citenamefont{Dahl, Mai, Cherf,
  Jetha, Garalde, Marziali, Akeson, Wang, and Lieberman}}]{Dahl2012}
\bibinfo{author}{\bibfnamefont{J.~M.} \bibnamefont{Dahl}},
  \bibinfo{author}{\bibfnamefont{A.~H.} \bibnamefont{Mai}},
  \bibinfo{author}{\bibfnamefont{G.~M.} \bibnamefont{Cherf}},
  \bibinfo{author}{\bibfnamefont{N.~N.} \bibnamefont{Jetha}},
  \bibinfo{author}{\bibfnamefont{D.~R.} \bibnamefont{Garalde}},
  \bibinfo{author}{\bibfnamefont{A.}~\bibnamefont{Marziali}},
  \bibinfo{author}{\bibfnamefont{M.}~\bibnamefont{Akeson}},
  \bibinfo{author}{\bibfnamefont{H.}~\bibnamefont{Wang}}, \bibnamefont{and}
  \bibinfo{author}{\bibfnamefont{K.~R.} \bibnamefont{Lieberman}},
  \bibinfo{journal}{Journal of Biological Chemistry}
  \textbf{\bibinfo{volume}{287}}, \bibinfo{pages}{13407}
  (\bibinfo{year}{2012}).

\bibitem[{\citenamefont{Lieberman et~al.}(2012)\citenamefont{Lieberman, Dahl,
  Mai, Akeson, and Wang}}]{Lieberman2012}
\bibinfo{author}{\bibfnamefont{K.~R.} \bibnamefont{Lieberman}},
  \bibinfo{author}{\bibfnamefont{J.~M.} \bibnamefont{Dahl}},
  \bibinfo{author}{\bibfnamefont{A.~H.} \bibnamefont{Mai}},
  \bibinfo{author}{\bibfnamefont{M.}~\bibnamefont{Akeson}}, \bibnamefont{and}
  \bibinfo{author}{\bibfnamefont{H.}~\bibnamefont{Wang}},
  \bibinfo{journal}{Journal of the American Chemical Society}
  \textbf{\bibinfo{volume}{134}}, \bibinfo{pages}{18816}
  (\bibinfo{year}{2012}).

\bibitem[{\citenamefont{Lieberman et~al.}(2013)\citenamefont{Lieberman, Dahl,
  Mai, Cox, Akeson, and Wang}}]{Lieberman2013}
\bibinfo{author}{\bibfnamefont{K.~R.} \bibnamefont{Lieberman}},
  \bibinfo{author}{\bibfnamefont{J.~M.} \bibnamefont{Dahl}},
  \bibinfo{author}{\bibfnamefont{A.~H.} \bibnamefont{Mai}},
  \bibinfo{author}{\bibfnamefont{A.}~\bibnamefont{Cox}},
  \bibinfo{author}{\bibfnamefont{M.}~\bibnamefont{Akeson}}, \bibnamefont{and}
  \bibinfo{author}{\bibfnamefont{H.}~\bibnamefont{Wang}},
  \bibinfo{journal}{Journal of the American Chemical Society}
  \textbf{\bibinfo{volume}{135}}, \bibinfo{pages}{9149} (\bibinfo{year}{2013}).

\bibitem[{\citenamefont{Lieberman et~al.}(2014)\citenamefont{Lieberman, Dahl,
  and Wang}}]{Lieberman2014}
\bibinfo{author}{\bibfnamefont{K.~R.} \bibnamefont{Lieberman}},
  \bibinfo{author}{\bibfnamefont{J.~M.} \bibnamefont{Dahl}}, \bibnamefont{and}
  \bibinfo{author}{\bibfnamefont{H.}~\bibnamefont{Wang}},
  \bibinfo{journal}{Journal of the American Chemical Society}
  \textbf{\bibinfo{volume}{136}}, \bibinfo{pages}{7117} (\bibinfo{year}{2014}).

\bibitem[{\citenamefont{Morin et~al.}(2015)\citenamefont{Morin, Cao,
  L{\'{a}}zaro, Arias-Gonzalez, Valpuesta, Carrascosa, Salas, and
  Ibarra}}]{Morin2015}
\bibinfo{author}{\bibfnamefont{J.~A.} \bibnamefont{Morin}},
  \bibinfo{author}{\bibfnamefont{F.~J.} \bibnamefont{Cao}},
  \bibinfo{author}{\bibfnamefont{J.~M.} \bibnamefont{L{\'{a}}zaro}},
  \bibinfo{author}{\bibfnamefont{J.~R.} \bibnamefont{Arias-Gonzalez}},
  \bibinfo{author}{\bibfnamefont{J.~M.} \bibnamefont{Valpuesta}},
  \bibinfo{author}{\bibfnamefont{J.~L.} \bibnamefont{Carrascosa}},
  \bibinfo{author}{\bibfnamefont{M.}~\bibnamefont{Salas}}, \bibnamefont{and}
  \bibinfo{author}{\bibfnamefont{B.}~\bibnamefont{Ibarra}},
  \bibinfo{journal}{Nucleic Acids Research} \textbf{\bibinfo{volume}{43}},
  \bibinfo{pages}{3643} (\bibinfo{year}{2015}).

\bibitem[{\citenamefont{Ren}(2016)}]{Ren2016}
\bibinfo{author}{\bibfnamefont{Z.}~\bibnamefont{Ren}},
  \bibinfo{journal}{Nucleic Acids Research} \textbf{\bibinfo{volume}{44}},
  \bibinfo{pages}{7457} (\bibinfo{year}{2016}).

\bibitem[{\citenamefont{Brenlla et~al.}(2014)\citenamefont{Brenlla, Markiewicz,
  Rueda, and Romano}}]{Brenlla2014}
\bibinfo{author}{\bibfnamefont{A.}~\bibnamefont{Brenlla}},
  \bibinfo{author}{\bibfnamefont{R.~P.} \bibnamefont{Markiewicz}},
  \bibinfo{author}{\bibfnamefont{D.}~\bibnamefont{Rueda}}, \bibnamefont{and}
  \bibinfo{author}{\bibfnamefont{L.~J.} \bibnamefont{Romano}},
  \bibinfo{journal}{Nucleic Acids Research} \textbf{\bibinfo{volume}{42}},
  \bibinfo{pages}{2555} (\bibinfo{year}{2014}).

\bibitem[{\citenamefont{SantaLucia and Hicks}(2004)}]{SantaLucia2004}
\bibinfo{author}{\bibfnamefont{J.}~\bibnamefont{SantaLucia}} \bibnamefont{and}
  \bibinfo{author}{\bibfnamefont{D.}~\bibnamefont{Hicks}},
  \bibinfo{journal}{Annual Review of Biophysics and Biomolecular Structure}
  \textbf{\bibinfo{volume}{33}}, \bibinfo{pages}{415} (\bibinfo{year}{2004}).

\end{thebibliography}
\end{document}